\documentclass[aps,prd,superscriptaddress,nofootinbib,amsmath,amsfonts,preprintnumbers,groupedaddress,showpacs,10pt,english]{revtex4-1}
\usepackage{amsmath}
\usepackage{amssymb}
\usepackage{babel}
\usepackage{wrapfig}
\usepackage{cancel}

\newcommand{\e}{\mathrm{e}}
\usepackage{relsize,exscale}
\makeatletter

\usepackage{array,multirow,graphicx}
\usepackage{dcolumn}
\usepackage{newlfont}
\usepackage{bm}
\usepackage[colorlinks,citecolor=blue,urlcolor=blue,linkcolor=blue]{hyperref}
\usepackage[figtopcap]{subfigure}
\usepackage{color}

\newcommand\be{\begin{equation}}
\newcommand\ba{\begin{eqnarray}}
\newcommand\ee{\end{equation}}
\newcommand\ea{\end{eqnarray}}

\begin{document}

\tolerance=5000
\allowdisplaybreaks[4]


\title{Black Holes with Electric and Magnetic Charges in $F(R)$ Gravity}

\author{G.~G.~L.~Nashed}
\email{nashed@bue.edu.eg}
\affiliation {Centre for Theoretical Physics, The British University, P.O. Box
43, El Sherouk City, Cairo 11837, Egypt}

\author{S.~Nojiri}
\email{nojiri@gravity.phys.nagoya-u.ac.jp}
\affiliation{Department of Physics, Nagoya University, Nagoya 464-8602, Japan \\
and \\
Kobayashi-Maskawa Institute for the Origin of Particles and the Universe, Nagoya University, Nagoya 464-8602, Japan}

\begin{abstract}
We construct spherically symmetric and static solutions in $F(R)$ gravity coupled with electromagnetic fields.
The solutions include new types of black holes with electric and magnetic charges.
We show that the higher-derivative terms make the curvature singularity much softer than that in the charged black holes in Einstein's general relativity.
We calculate some thermodynamical quantities of the obtained black holes like entropy, Hawking radiation, and quasi-local energy and
we confirm that the black hole solutions satisfy the first law of thermodynamics.
Finally, we study the stability analysis using the odd-type mode and show that there are stable black hole solutions and the radial speed of the parity-odd mode
is unit, that is, the speed of light.
\end{abstract}


\maketitle
\section{Introduction}

For a few decades, $F(R)$ gravity theory has attracted much attention because the theory may explain the accelerating expansion of the present universe in addition to
the inflation of the early universe \cite{Capozziello:2002rd,Capozziello:2003gx,Carroll:2003wy,Nojiri:2003ft,Hu:2007nk,Starobinsky:2007hu}
(for reviews, see \cite{Nojiri:2006ri,Copeland:2006wr,Sotiriou:2008rp,Nojiri:2010wj,Capozziello:2011et,Nojiri:2017ncd}).
Furthermore, the possibility that the dark matter might be explained by the $F(R)$ gravity has been investigated \cite{Nojiri:2006ri, Copeland:2006wr, Nojiri:2010wj, Clifton:2011jh}.
The existence of higher-order curvature terms in general relativity also supplies interesting physical results, for example, it makes the condensation harder to be formed
in holographic superconductivity \cite{Gregory:2009fj, Kuang:2013oqa}, it also amends the low-energy tensor perturbation spectrum
in string backgrounds \cite{Gasperini:1997up}, and it affects the dynamics of stellar structure \cite{Hansraj:2020xmz}.

In the $F(R)$ gravity theory, the scalar curvature $R$ in the Einstein-Hilbert action is replaced by an adequate function $F(R)$ of $R$ \cite{Capozziello:2002rd,Capozziello:2003gx,Nojiri:2003ft,
Cognola:2007zu,Pogosian:2007sw,Zhang:2005vt,Li:2007xn,Song:2007da,Nojiri:2007cq,Nojiri:2007as,Capozziello:2018ddp,Vainio:2016qas}.
We may consider the higher derivative theories including the Ricci or Riemann curvatures in the actions not only in the form of the scalar curvature.
Such higher-order corrections to the action of the general relativity yield a renormalizable and therefore quantizable theory of gravity \cite{Stelle:1976gc}.
We should note, however, that such higher-derivative theories except the $F(R)$ gravity have the Ostrogradski instability \cite{Ostrogradsky:1850fid} which
is problematic because there appear ghosts, which make the theory inconsistent \cite{Woodard:2006nt}.

To investigate whether the $F(R)$ theory is realistic, spherically symmetric and static black hole solutions have been investigated
\cite{Multamaki:2006zb,delaCruz-Dombriz:2009pzc,Hendi:2011hxq,Nashed:2021sey,Nashed:2021mpz,Nashed:2021lzq,
Nashed:2021ffk,Nashed:2020mnp,Nashed:2020kdb,Nashed:2020tbp,Tang:2019qiy,Nashed:2018oaf,Nashed:2018efg,Nashed:2018piz}.
In the gravitational collapse, all the matters including the charged ones are absorbed into the black hole, and therefore even in the $F(R)$ gravity, there must exist
charged black hole.
Thus, it is important to investigate the stationary black hole solutions in the $F(R)$ gravity coupled with the electromagnetic fields and the contribution from the electromagnetic
fields to the geometry in the framework of the $F(R)$ gravity.
The black hole solutions in $F(R)$ gravity in vacuum or in the case with electromagnetic fields have been studied in
\cite{Multamaki:2006zb,Nashed:2020kdb,Sebastiani:2010kv,Hendi:2014mba,Multamaki:2006ym,
Nashed:2020tbp,Nashed:2019uyi,Nashed:2019tuk,delaCruz-Dombriz:2009pzc,Jaryal:2021lsu,Eiroa:2020dip}
and in \cite{Tang:2020sjs, Karakasis:2021lnq, Karakasis:2021rpn}, the scalar fields have been included as matter in three and four dimensions
while the $F(R)$ gravity coupled with/without non-minimally coupled scalar fields as a matter has been studied in the frame of
cosmology \cite{Pi:2017gih,delaCruz-Dombriz:2016bjj}.
Capozziello et al. have used the Noether symmetry to investigate the spherically symmetric solutions \cite{Capozziello:2007wc, Capozziello:2012iea}
and for the axially symmetric black hole solution \cite{Capozziello:2009jg}.
Dynamical spherically symmetric black hole solutions have been also presented in \cite{Elizalde:2020icc,Nashed:2019yto,Nashed:2019tuk} for a specific form of $F(R)$.
As the topics related to the strong gravitational background, not only the static spherically symmetric
black holes \cite{Sultana:2018fkw,Canate:2017bao,Yu:2017uyd,Canate:2015dda,Kehagias:2015ata,Nelson:2010ig,delaCruz-Dombriz:2009pzc}
but neutron star have been investigated \cite{Feng:2017hje,AparicioResco:2016xcm,Capozziello:2015yza,Staykov:2018hhc,Doneva:2016xmf,Yazadjiev:2016pcb,Yazadjiev:2015zia,Yazadjiev:2014cza,
Ganguly:2013taa,Astashenok:2013vza,Orellana:2013gn,Arapoglu:2010rz,Cooney:2009rr} in  the form $F(R)=R+\alpha R^2$.
It is also well-known that the $F(R)$ gravity can be rewritten as the Brans-Dicke theories \cite{Brans:1961sx} that have a scalar potential of
the gravitational origin \cite{Chiba:2003ir,OHanlon:1972xqa,Chakraborty:2016gpg,Chakraborty:2016ydo}.
In this paper, we construct a new type of spherically symmetric and static black hole with electric and magnetic charge in
the framework of the $F(R)$ gravity and investigate the physical properties of such black hole solutions.

In Sec.~\ref{S2}, we review the fundamentals of $F(R)$ gravity, and in Sec.~\ref{S3}, we apply the field equations in the $F(R)$ gravity
to a spherically symmetric space-time.
There appears a system of differential equations that has three unknown functions and
we derive the solutions of this system that is characterized by a convolution function.
If this convolution function vanishes, we obtain the black hole solution in general relativity, that is, the Schwarzschild solution.
Hence the convolution function could appear due to higher-order curvature terms that characterize the $F(R)$ gravity.
By calculating the Kretschmann scalar, the Ricci tensor square, and the Ricci scalar, we show that the singularities
in such invariants become weaker than those of general relativity black holes.
In Sec.~\ref{S4}, we calculate the thermodynamical quantities in the obtained black hole solutions to compare them with those in the known solutions.
In Sec.~\ref{S616}, we use the odd-type method and study the stability of these black hole solutions.
In the final section, we give our concluding remarks.

\section{Basic equations in $F(R)$ gravity coupled with electromagnetic fields}\label{S2}

We may regard the $F(R)$ gravity as an extension of general relativity that was investigated in
\cite{Buchdahl:1970ynr,Capozziello:2011et,Nojiri:2010wj,Nojiri:2017ncd,Capozziello:2003gx,Capozziello:2002rd,Nojiri:2003ft,Carroll:2003wy}.
The action of the $F(R)$ gravity coupled with the electromagnetic fields is given by
\begin{align}
\label{a1}
S=S_\mathrm{g}+S_\mathrm{EM}\,,
\end{align}
where $S_\mathrm{g}$ is given by
\begin{align}
\label{a2}
S_\mathrm{g}=\frac{1}{2\kappa^2} \int d^4x \sqrt{-g} F(R)\, .
\end{align}
Here $\kappa$ is the gravitational constant, $R$ is the Ricci scalar, $g$ is the determinant of the metric, and $F(R)$ is an analytic function of $R$.
On the other hand, $S_\mathrm{EM}$ is given by
\begin{align}
\label{a3}
S_\mathrm{EM}=-\frac{1}{2}\int d^4x \sqrt{-g} F_{\mu \nu}F^{\mu\nu}\,,
\end{align}
where $F_{\mu \nu} =\partial_\mu A_\nu - \partial_\nu A_\mu$ and $A_\mu$ is the gauge potential.

In the following, we choose the unit where $\kappa=1$.
The variations of the action (\ref{a1}) with respect to the metric tensor $g_{\mu \nu}$ and the gauge potential $A_\mu$ give
the field equations as in \cite{Cognola:2005de,Koivisto:2005yc},
\begin{align}
\label{f1}
0=&\, R_{\mu \nu} F_R-\frac{1}{2}g_{\mu \nu}F +g_{\mu \nu} \Box F_R-\nabla_\mu \nabla_\nu F_R-8\pi T_{\mu \nu}\,, \\
\label{fe2}
0=&\, \partial_\nu \left( \sqrt{-g} {F}^{\mu \nu} \right)\,.
\end{align}
Here $R_{\mu \nu}$ is the Ricci tensor\footnote{The Ricci tensor is defined as
\[
R_{\mu \nu}=R^{\rho}_{\ \mu \rho \nu}= \Gamma^\rho_{\mu \nu,\rho} - \Gamma^\rho_{\mu\rho, \nu}
+ \Gamma^\rho_{\beta \rho}\Gamma^\beta_{\nu\mu} - \Gamma^\rho_{\beta\nu}\Gamma^\beta_{\rho\mu}\, ,
\]
with e $\Gamma^\rho_{\mu \nu}$ being the Christoffel second kind symbols.}
and $\Box$ is the d'Alembertian operator which is defined as
$\Box= \nabla_\alpha\nabla^\alpha$ where $\nabla_\alpha B^\alpha$ means the covariant differentiation of the vector $B^\alpha$ and $F_R=\frac{dF(R)}{dR}$.
The energy-momentum tensor of the electromagnetic fields $T_{\mu \nu}$ is defined as
\begin{align}
T_{\mu \nu}:=\frac{1}{4\pi}\left(g_{\rho\sigma}{{F}_\nu^{\ \rho}}F_\mu^{\ \sigma}-\frac{1}{4} g_{\mu \nu} F^2\right)\, .
\end{align}
The trace of the field equations (\ref{f1}) has the following form,
\begin{align}
\label{f3333}
0=RF_R-2F+3\Box F_R \,.
\end{align}
By rewriting Eq.~(\ref{f3333}) as $F=\frac{1}{2}\left(RF_R+3\Box F_R \right)$ and deleting $F$ in Eq.~(\ref{f1}), we obtain
\begin{align}
\label{f3ss}
0=R_{\mu \nu} \left[ 1+F_R \right] - \frac{1}{4}g_{\mu \nu}R \left[1+F_R \right] + \frac{1}{4}g_{\mu \nu}\Box F_R
 -\nabla_\mu \nabla_\nu F_R-8\pi T_{\mu \nu} \,.
\end{align}
In the next section, we apply the field equations~(\ref{fe2}) and (\ref{f3ss}) to spherically symmetric and static space-time and derive exact solutions describing the black hole
with electric and/or magnetic charges.

\section{Spherically symmetric and static solutions}\label{S3}
In this section, we do not specify the function $F(R)$ as a function of $R$ but we assume the radial coordinate $r$ dependence of $F_R$.
After solving the equations, we consider the functional form of $F(R)$.
This tells that there exists a function $F(R)$ which generates the solution.

We first assume the following spherically symmetric and static space-time with two functions $h(r)$ and $h(r)$ of the radial coordinate $r$,
\begin{align}
\label{met12}
ds^2=-h(r)dt^2+\frac{dr^2}{h_1(r)}+r^2 \left( d\theta^2+\sin^2\theta d\phi^2 \right)\, .
\end{align}
The metric (\ref{met12}) gives the following Ricci scalar,
\begin{align}
\label{Ricci}
R(r)=\frac{r^2h_1h'^2-r^2hh'h'_1-2r^2hh_1h''-4rh[h_1h'-hh'_1]+4h^2(1-h_1)}{2r^2h^2}\,,
\end{align}
where $h\equiv h(r)$, $h_1\equiv h_1(r)$, $h'=\frac{dh}{dr}$, $h''=\frac{d^2h}{dr^2}$ and $h_1'=\frac{dh}{dr}$.
For the metric (\ref{met12}), by using Eq.~(\ref{Ricci}), we find Eqs.~(\ref{fe2}) and (\ref{f3ss}) have the following forms, \\
The $tt$, $t\phi$, $rr$, $r\theta$, $r\phi$, $\theta\theta$, and $\phi\phi$ components of Eq.~(\ref{f3ss}) are given by
\begin{align}
\label{tt}
0=&\, \frac{1}{8h^2r^4 \sin^2 \theta }\left\{2 h^2r^4h_1 \sin^2 \theta F''_1-2 F_1 r^4h_1 h \sin^2 \theta h'' -F_1r^4h_1 \sin^2\theta h'^2
+ \left( 3 F'_1h_1 r+F_1 \left( 4 h_1 +h'_1 r \right) \right) h r^3 \sin^2\theta h' \right. \nonumber\\
&\, +h \left[ r^3h \sin^2\theta \left( 4 h_1 + h'_1 r \right) F_1+4 h F_1 r^3 \sin^2\theta h'_1 +8 h_1 h r^2 n_\phi^2-16 h_1 h r^2 n_\phi p'
+8 h_1 h r^2 p'^2+ 8 r^4 \sin^2 \theta h_1 q'^2 \right. \nonumber\\
&\, \left. \left. +4 h \left( 2 k_\theta^2+F_1 r^2 \sin^2\theta \left(h_1 -1 \right) \right) \right] \right\} \, , \\
\label{t ph}
0=&\, 2h_1q' \left( n_\phi-p' \right) \,, \\
\label{rr}
0=&\,\frac{1}{8h^2r^4 \sin^2 \theta } \left\{F_1 r^4h_1\sin^2\theta h'^2-6 h^2r^4h_1\sin^2\theta F''_1 -2 F_1 r^4h_1 h \sin^2\theta h''
+ \left( F'_1 h_1 r-F_1 \left( h'_1 r -4 h_1 \right) \right) h r^3\sin^2\theta h' \right. \nonumber\\
&\, + \left[ \left(4h_1- 3h'_1 r \right) h r^3\sin^2\theta F'_1 -4h F_1 r^3\sin^2\theta h'_1 -8h_1 h r^2 n_\phi^2+16h_1 h r^2 n_\phi p' -8h_1 h r^2p'^2
+8r^4h_1\sin^2\theta q'^2 \right. \nonumber\\
& \,\left. \left. +4h \left( 2 k_\theta^2+F_1 r^2\sin^2\theta \left( h_1-1 \right) \right) \right] h \right\}\,, \\
\label{th ph}
0=&\, 2k_\theta \left( n_\phi-p' \right)\,, \\
\label{ph th}
0=&\, 2h_1k_\theta \left( n_\phi-p' \right)\,, \\
\label{th th}
0=&\, \frac{1}{8h^2r^4 \sin^2 \theta } \left\{2 h^2r^4h_1\sin^2\theta F''_1 +2 F_1 r^4h_1 h \sin^2\theta h''
+ \left[ \left( h'_1 r -4 h_1 \right) h +h_1 h' r \right] h r^{3} \sin^2\theta F'_1 \right. \nonumber\\
&\, -F_1 r^4h_1 \sin^2\theta h'^2 +F_1 r^4 h'_1 h \sin^2\theta h' \nonumber \\
&\, \left. +4 \left( 2h_1 h r^2 n_\phi^2-4 h_1 h r^2n_\phi p' +2h_1 h r^2 p'^2-2r^4h_1\sin^2\theta q'^2-h \left( 2 k_\theta^2+F_1 r^2\sin^2\theta \left(h_1-1 \right) \right) \right) h
\right\}\,, \\
\label{ph ph}
0=&\, \frac{1}{8h^2r^4 \sin^2 \theta }\left\{2 h^2r^4h_1\sin^2\theta F''_1 +2 F_1 r^4h_1 h \sin^2\theta h'' + \left[ \left( h'_1 r -4 h_1 \right) h
+h_1 h' r \right] h r^{3} \sin^2\theta F'_1 \right. \nonumber\\
&\, -F_1 r^4h_1 \sin^2\theta h'^2 +F_1 r^4 h'_1 h \sin^2\theta h' \nonumber \\
&\, \left. -4 \left( 2h_1 h r^2 n_\phi^2-4 h_1 h r^2n_\phi p' +2h_1 h r^2 p'^2+2r^4h_1\sin^2\theta q'^2+h \left( 2 k_\theta^2+F_1 r^2\sin^2\theta \left(h_1-1 \right) \right) \right) h
\right\}\,.
\end{align}
The trace-component (\ref{f3333}) has the following form,
\begin{align}\label{tra}
0=&\, \frac{1}{h^2r^2} \left\{6 h^2r^2 F''_1 h_1 -2 F_1 h_1 h'' h r^2+F_1 h_1 h'^2r^{2 }+ \left[ 3 F'_1 h_1 r-F_1 \left( 4 h_1 + h'_1 r \right) \right] h rh' \right. \nonumber \\
&\, \left. + h^2 \left[ \left(3r^2h'_1 +12 h_1 r \right) F'_1 -4rF_1 h'_1-4\left( h_1-1 \right) F_1 -4F r^2 \right] \right\}\,.
\end{align}
The non-vanishing components of the field equations (\ref{fe2}) are $t$, $r$, and $\phi$ components which have the following forms, respectively,
\begin{align}
\label{t}
0=&\, rhh'_1q'-rh_1h'q'+2rhh_1q''+4hh_1q'\,, \\
\label{tt2}
0=&\, n_{\phi \phi}\,, \\
\label{tt3}
0=&\, h_1 h' r^2\sin \theta n_\phi -h_1 h'r^2\sin \theta p' +r^2h \sin \theta) h'_1 n_\phi -r^2A \sin \theta h'_1p' \nonumber \\
&\, -2 r^2h \sin \theta h_1 p'' +2 h k_\theta \cos \theta -2 h k_{\theta \theta} \sin \theta\,.
\end{align}
For brevity, we put $F_1\equiv F_1(r)=\frac{dF\left(R \left(r\right) \right)}{dR(r)}$, $F'_1=\frac{dF_1(r)}{dr}$, $F''_1=\frac{d^2F_1(r)}{dr^2}$, $n_\phi=\frac{dn(\phi)}{d\phi}$,
$k_\theta=\frac{dk(\theta)}{d\theta}$, and $q'=\frac{dq(r)}{dr}$.
Here $q(r)$, $k(\theta)$, $p(r)$, and $n(\phi)$ are the components of the electric and magnetic field components defined as
\begin{align}
\label{p}
A_\mu dx^\mu :=q(r)dt+ n(\phi)dr+ \left[ p(r)+k(\theta) \right]d\phi\, .
\end{align}
We stress that when the magnetic field vanishes, i.e., $n=p=k=0$, the components of the field equations (\ref{f3ss}) for $\theta\theta$
and $\phi\phi$ components become identical with each other and in that case, the field equations (\ref{tt}), (\ref{t ph}), (\ref{rr}), (\ref{th ph}), (\ref{ph th}), (\ref{th th}), (\ref{ph ph}),
and (\ref{tra}) coincide with those derived in \cite{Nashed:2019tuk}.

Now we discuss two cases that $h=h_1$ and that $h\neq h_1$\footnote{Because the present study deals with spherically symmetric case,
we assume $F(R)$ only depends on the radial coordinate $r$, $F(R)=F(r)$.}.
Because we do not assume the specific form of $F(R)$ as a function of $R$, we have seven unknown functions, $h$, $h_1$, $q$, $n$, $k$, $p$, and $F_1=\frac{dF(R)}{dR}$
but we have six independent differential equations.
To solve these differential equations, we assume that $F_1$ is given by
\begin{align}
\label{ass1n}
F_1=1+\frac{a}{r^3}\,.
\end{align}
Here $a$ is a constant.
Eq.~(\ref{ass1n}) shows that when $a=0$, we return to the case of general relativity where $F_1(R)=\mathrm{const}.$

\ \\

\noindent
\underline{The case $h=h_1$}

In the case of $h=h_1$, the field equations (\ref{th ph}), (\ref{ph th}), (\ref{th th}), (\ref{ph ph}) and (\ref{tra}) have no solution whenever $a\neq 0$.
This means that when $h=h_1$ and $a\neq0$, we will not obtain any solution, and when $a=0$, we obtain the following solution,
\begin{align}
\label{general relativity}
h=r^2c_0+1+\frac{c_1}r+\frac{{c_2}^2+{c_3}^2}{r^2}\,, \quad q(r)=\frac{c_2}r\,, \quad k(\theta)=c_3\cos\theta\,,\quad p(r)=c_4r\,, \quad n(\phi)=c_5\phi\,.
\end{align}
Here $c_0$, $c_1$, $c_2$, $c_3$, and $c_4$ are the constants of the integration.
The above discussion is consistent with the previous studies which ensure that any solution with $h=h_1$ in the frame of $F(R)$ will not differ
from the black hole in the general relativity and $F(R)$ only play the role of a cosmological constant \cite{Multamaki:2006zb, Nashed:2018oaf, Nashed:2018efg,
Nashed:2018piz}.
The above solution (\ref{general relativity}) supports this discussion.
The invariant scalars of the above solution (\ref{general relativity}) have the following form,
\begin{align}
\label{R1}
&K=R_{a b c d}R^{a b c d}=48 {c_0}^2+\frac{12{c_1}^2}{r^6}+\frac{48c_1({c_2}^2+{c_3}^2)}{r^7}+\frac{56({c_2}^2+{c_3}^2)^2}{r^8}\,,\nonumber\\
&R_{a b}R^{a b}=R=36 {c_0}^2+\frac{4({c_2}^2+{c_3}^2)^2}{r^8}\,, \quad R= -12 c_0\,.
\end{align}
The above invariants show there are the contribution of the electric charge as well as the magnetic field.
Now we are going to study the case $h\neq h_1$.

\ \\

\noindent
\underline{The case $h\neq h_1$}

In this case, the system of differential equations can be solved analytically and the solution has the following form,
\begin{align}
\label{fr1}
h=256c_6h_1(r) \left(1-\frac a{2r^3}\right)^8\, .
\end{align}
where $h_1$ is given by \begin{align}
\label{frB1}
h_1 =&\, \frac{r^{14} \left( -972 \ln \left( r \right) r^{12}
+324 \ln \left( a+r^3 \right) r^{12}-260\,a r^9+ a^4+66 a^2r^6
-12 a^3r^3 \right) c_7 }{\left( a-2 r^3 \right)^8}
+{\frac{r^{26}c_8}{\left( a-2 r^3 \right)^8}}\nonumber\\
&\, -\frac{1390932}{455 a^\frac{4}{3}{c_6} \left( a-2 r^3 \right)^8}
\left\{\sqrt{3} \left({c_2}^2c_6 -{\frac{507 a^\frac{2}{3}c_6}{2332}}
+207025{c_3}^2\right) r^{26} \arctan \left( {\frac{\sqrt[3] a-2r }{\sqrt{3}\sqrt[3] a}} \right) \right. \nonumber \\
&\, -\frac{r^{26}}{2} f_1 \ln \left( r^2-r\sqrt[3] a+ a^\frac{2}{3} \right)
+f_1 r^{26}\ln \left( r+\sqrt[3] a \right) \nonumber\\
&\, + \frac{2035345r^{16} a^{10/3}}{122402016}
\left( \frac{56207184}{10176725}{c_2}^2c_6+r^2c_6+ \frac{35803976208}{31313} {c_3}^2\right)
 - \frac{1843153}{168302772} r^{13} \left( {\frac{57119260}{23960989}}{c_2}^2c_6 \right. \nonumber\\
&\, \left. +\frac{69971093500}{141781}{c_3}^2+r^2c_6 \right) a^{13/3}
+{\frac{1196689r^{10} a^{16/3}}{214203528}} \left( \frac{189343}{184106}{c_2}^2c_6
+ \frac{3015287275}{14162}{c_3}^2+r^2c_6 \right) \nonumber \\
&\, -\frac{41171}{23645844} \left( {\frac{407388} {782249}}{c_2}^2c_6 +r^2c_6+{\frac{6487653900}{60173}}{c_3}^2 \right)
r^7 a^{\frac{19}{3}}\nonumber\\
&\, +\frac{20189}{61201008} \left( \frac{86504600}{1553}{c_3}^2+ \frac{5432}{20189}{c_2}^2c_6+r^2c_6 \right) r^4 a^{\frac{22}{3}}
 -\frac{12805}{351905796} \left( \frac{7084}{64025}{c_2}^2c_6+r^2c_6 \right. \nonumber\\
&\, \left. + \frac{4512508}{197}{c_3}^2 \right) r a^{\frac{25}{3}}
+ \frac{7416721}{30600504} \left( \frac{324676001650}{570517}{c_3}^2+r^2c_6+ \frac{20387818}{7416721} {c_2}^2c_6 \right) r^{22} a^\frac{4}{3}
+\frac{5}{2781864} a^{\frac{28}{3}}c_6\nonumber\\
&\, \left. -3 r^{25}\sqrt[3] a \left( 207025 {c_3}^2+{c_2}^2c_6 \right) -\frac{3572881}{76501260}
\left( \frac{139814180}{25010167}{c_2}^2c_6+\frac{318077259500}{274837}{c_3}^2+r^2c_6 \right) r^{19} a^{7/3} \right\} \,,
\end{align}
and , $q$, $p$, $k$, and $n$ are given by
\begin{align}
\label{qpkn}
q(r)=&\, \frac{c_2}r\,, \quad p(r)=c_4r\,, \quad k(\theta)=c_3\cos\theta\,,\quad n(\phi)=c_5\phi\, .
\end{align}
{Here $c_2$, $c_3$, $c_4$, $c_5$, $c_6$, and $c_7$  are the constants of the integration, again,
and $f_1$ is a constant that is defined as $f_1= {c_2}^2c_6+ \frac{507 a^\frac{2}{3}c_6}{2332} +207025{c_3}^2$.
In spite that Eq.~(\ref{fr1}) seems to tell that we could obtain $h\propto h_1$ when $a=0$, this is not true because
the third term of $h_1$ in (\ref{frB1}) includes the inverse power of the dimensional parameter $a$ and therefore $a$ cannot vanish.}

By substituting Eqs.~(\ref{fr1}), (\ref{frB1}), and (\ref{qpkn}) into the trace equation~(\ref{tra}), we obtain a very lengthy expression $f(r)$ whose asymptotic form up to
$\mathcal{O}\left(\frac{1}{r^6}\right)$ is given by
\begin{align}
\label{f1(r)}
F(r)=&\, \frac{1043199\pi \sqrt{3}}{14560a^{1/3}r^3} \left( {c_2}^2+52998400 {c_3}^2- \frac{507}{2332} a^\frac{2}{3} \right) \pi \nonumber \\
&\, -\frac{1}{5r^5}\left\{\frac{9388791\sqrt{3}}{14560a^{1/3} L}\left( {c_2}^2+52998400{c_3}^2- \frac{507}{2332} a^\frac{2}{3} \right) \pi \right. \nonumber\\
&\, \left. + {\frac{-15300252 \pi \sqrt{3} \left( {c_2}^2+52998400 {c_3}^2 \right) a^\frac{2}{3}-2038400 a^2L+3326427 a^\frac{4}{3}\sqrt{3}\pi }{29120a L}} \right\}+ c_9 \,,
\end{align}
where $L$ is a constant defined as $L=\frac{10010 c_8 a^\frac{4}{3}-810888875596800 \sqrt{3}\pi {c_3}^2
 -15300252 \sqrt{3} \pi {c_2}^2+3326427 \sqrt{3}\pi a^\frac{2}{3}}{2562560a^\frac{4}{3}}$.
Using Eqs.~(\ref{fr1}), (\ref{frB1}), and (\ref{qpkn}), we obtain a lengthy form of the Ricci scalar in (\ref{Ricci}) whose asymptotic form up to $\mathcal{O}\left(\frac{1}{r^6}\right)$ is given by
\begin{align}
\label{ric}
R\approx&\, -{\frac{1}{640640L\sqrt[3] ar^5}}\left\{7687680\, L^2\sqrt[3] ar^5-2432666626790400\,\sqrt{3}\pi \,Lr^2{c_3}^2-45900756
\sqrt{3}\pi \,Lr^2{c_2}^2 \right. \nonumber \\
&\, \left. +9979281 \sqrt{3}\pi Lr^2 a^\frac{2}{3}-8968960 a^\frac{4}{3}C+810888875596800 \sqrt{3}\pi {c_3}^2+15300252 \sqrt{3}\pi {c_2}^2
 -3326427\sqrt{3}\pi a^\frac{2}{3}\right\}\,.
\end{align}
From Eq.~(\ref{ric}), we obtain $r(R)$ as
\begin{align}
\label{ric1}
r\approx \frac{27 \sqrt[3]{33647995180800}}{40040 R+480480 L}\sqrt[3]{\sqrt{3} \left(\frac{1}{12} R+L \right)^2\pi
\left({c_2}^2+52998400 {c_3}^2- \frac{507}{2332} a^\frac{2}{3} \right) \frac{1}{\sqrt[3] a}} \,.
\end{align}
Using Eq.~(\ref{ric1}) in Eq.~(\ref{f1(r)}), we obtain the following expression of $F(R)$
\begin{align}
\label{f2(r)}
F(R)=C_0+C_1R+C_2R^2+C_3R^3\cdots\,,
\end{align}
where $C_i$, $i=0,1, \cdots$ are constants defined as
\begin{align}
\label{f2(r)B}
C_0=&\, {\frac{{2332}^\frac{2}{3}\sqrt[3]{33647995180800}}{2803999598400}} \nonumber \\
&\, \times \left\{{14428814400}^\frac{2}{3}
\left( {\frac{\pi \sqrt{3} \left(507 a^\frac{2}{3} -2332{c_2}^2-123592268800{c_3}^2 \right) L^2}{\sqrt[3] a}} \right)^\frac{5}{3}
\left( \frac{c_9}{12}+L \right) \sqrt[3] a \right. \nonumber\\
&\, \left. -{\frac{4335057126400}{81}} L^4 a^\frac{2}{3}\sqrt{3}\pi+{\frac{230064816357376000}{1594323}} L^4 \left( {\frac{347733}
{203840}} \pi \sqrt{3} \left( {c_2}^2+52998400 {c_3}^2 \right) +L a^\frac{4}{3} \right) \right\} \nonumber\\
&\, \times \left( 2332 \sqrt{3} \left({
\frac{507}{2332}} a^\frac{2}{3}- {c_2}^2-52998400 {c_3}^2 \right) {\frac{L^2\pi}{\sqrt[3]{a}}} \right)^{-\frac{5}{3}}{\frac{1}{\sqrt[3] a}}
\,,\nonumber\\
C_1=&\, -\frac{16307200}{2281476213}{2332}^\frac{2}{3}\sqrt[3]{33647995180800}
\left\{\frac{255879}{689920} L^3 a^\frac{2}{3}\sqrt{3}\pi -\frac{4782969}{1150324081786880000} \sqrt[3]{(14428814400)^2} \right. \nonumber\\
&\, \times \left( \frac{\sqrt{3}\pi \left( -123592268800 {c_3}^2-2332 {c_2}^2+507 a^\frac{2}{3} \right) L^2}{\sqrt[3] a} \right)^\frac{5}{3}\sqrt[3] a \nonumber \\
&\, \left. + \left( -\frac{347733}{203840} \pi \left(52998400 {c_3}^2+{c_2}^2 \right) \sqrt{3}+L a^\frac{4}{3} \right) L^3 \right\} \nonumber\\
& \times \left( -2332 \sqrt{3} \left( 52998400 {c_3}^2+{c_2}^2-\frac{507}{2332} a^\frac{2}{3} \right) {\frac{\pi L^2}{\sqrt[3] a}} \right)^{-\frac{5}{3}}{\frac{1}{\sqrt[3] a}}\,,\nonumber\\
C_2=&\, -{\frac{4076800}{20533285917}} \left( {\frac{255879}{689920}}\, a^\frac{2}{3}\sqrt{3}\pi -{\frac{347733}{203840}} \pi \left( 52998400 {c_3}^2
+{c_2}^2 \right) \sqrt{3}+L a^\frac{4}{3} \right) {2332}^\frac{2}{3}\sqrt[3]{33647995180800} L^2\nonumber\\
&\, \times \left( -2332 \sqrt{3} \left( 52998400 {c_3}^2+{c_2}^2-{\frac{507}{2332}} a^\frac{2}{3} \right) {\frac{\pi C^2}{\sqrt[3] a}}
\right)^{-\frac{5}{3}}{\frac{1}{\sqrt[3] a}}\,,\nonumber\\
C_3=&\, {\frac{1019200}{554398719759}} \left( {\frac{255879}{689920}} a^\frac{2}{3}\sqrt{3}\pi -{\frac{347733}{203840}} \pi
\left( 52998400 {c_3}^2+{c_2}^2 \right) \sqrt{3}+L a^\frac{4}{3} \right) {2332}^\frac{2}{3}\sqrt[3]{33647995180800}L\nonumber\\
&\, \times \left( -2332 \sqrt{3} \left( 52998400 {c_3}^2+{c_2}^2-{\frac{507}{2332}} c^\frac{2}{3} \right) {\frac{\pi C^2}{\sqrt[3] a}} \right)^{-\frac{5}{3}}{\frac{1}{\sqrt[3] a}}\,.
\end{align}
{Eqs.~(\ref{fr1}), (\ref{frB1}), (\ref{qpkn}), (\ref{f1(r)}), and (\ref{ric}) show that the dimensional parameter $a$ cannot vanish.
Thus we have a new charged black hole solution that does not coincide with any charged black hole solution of general relativity.
When the constants $c_2$ and $c_3$ vanish, however, we can put the dimensional parameter $a$ to vanish,
and in this case, we obtain the Schwarzschild black hole of general relativity, i.e.,}
\begin{align}
\label{reda}
h(r)=h_1(r)=r^2c_0+1+\frac{c_1}r \quad \mbox{and} \quad F(r)=1\, .
\end{align}
We should note that if the form of $F(R)$ is given as in (\ref{f2(r)}) with (\ref{f2(r)B}), it is difficult to find the general solution.
It is clear, however, that there should really exist a model realizing the black hole in (\ref{fr1}), (\ref{frB1}), and (\ref{qpkn}).

In the next section, we investigate the physical properties of the black hole given by (\ref{fr1}), (\ref{frB1}), and (\ref{qpkn}).

\subsection{The physical properties of the black hole (\ref{general relativity}) and the black hole (\ref{fr1}), (\ref{frB1}), and (\ref{qpkn})}

Before clarifying the properties of the black hole given by (\ref{fr1}), (\ref{frB1}), and (\ref{qpkn})),
we investigate the physical properties of the black hole in (\ref{general relativity}).
For the purpose, we rewrite $h(r)=h_1(r)$ in Eq.~(\ref{general relativity}) as follows,
\begin{align}
\label{mpab}
h(r)=\Lambda r^2+ 1-\frac{2M}r-\frac{{q_1}^2+{q_2}^2}{r^2}\,,
\end{align}
where we have written the constants as $c_0=\Lambda$, $c_1=-2M$, $c_2=q_1$, and $c_3=q_2$ in Eq.~(\ref{general relativity}).
By using Eq.~(\ref{mpab}), the metric in (\ref{met12}) is given by
\begin{align}
\label{metaf}
ds^2=-\left[\Lambda r^2+ 1-\frac{2M}r-\frac{{q_1}^2+{q_2}^2}{r^2}\right]dt^2
+\frac{dr^2}{\Lambda r^2+ 1-\frac{2M}r-\frac{{q_1}^2+{q_2}^2}{r^2}}+r^2 \left( d\theta^2+\sin^2d\phi^2 \right)\,.
\end{align}
Therefore the space-time is asymptotically AdS/dS space-time and the metric coincides with that of the Reissner-Nordstr{\" o}m space-time when $q_2=0$,
which tells that $q_2$ is the contribution that comes from the magnetic field.

We now discuss the properties of the black hole in Eqs.~(\ref{fr1}), (\ref{frB1}), and (\ref{qpkn}).
We rewrite the asymptotic form when $r$ is large, $h(r)$ and $h_1(r)$ in the solution (\ref{fr1}), (\ref{frB1}), and (\ref{qpkn}) as follows,
\begin{align}
\label{mpabB}
h(r)\approx&\, \Lambda r^2+ 1-\frac{2M}r+\frac{{q_3}^2+{q_4}^2}{r^2}-\frac a{r^3}-\frac{2a \left( {q_4}^2+{q_3}^2
\right) }{r^5}+\frac{31a^2}{40r^6}+\frac{87}{35} \frac{a^2 \left( {q_4}^2+{q_3}^2 \right) }{r^8}-\frac{127a^3}{220r^9}\nonumber\\
&\, -\frac{a^\frac{2}{3} \left(\sqrt{316437030} \sqrt{3}\pi {q_4}^2+2562560 a^\frac{4}{3}\Lambda-3326427 a^\frac{2}{3}
\sqrt{3}\pi +12812800 \sqrt[3] aM+15300252 \sqrt{3}\pi {q_2}^2 \right) }{2562560r^4}\nonumber\\
&\, - \frac{3a^\frac{5}{3} \left( 15300252 \sqrt{3}\pi {q_3}^2+\sqrt{15300252} \sqrt{3}\pi {q_4}^2
+7687680 \sqrt[3] aM+2562560 a^\frac{4}{3}\Lambda-3326427 a^\frac{2}{3}\sqrt{3} \pi \right) }{2562560r^7} \,,\nonumber\\
h_1(r)\approx&\, \Lambda r^2+ 1-\frac{1281280 \sqrt[3] aM-\sqrt{3}\pi \left[3326427 a^\frac{2}{3}
 -\sqrt{15300252} {q_4}^2 -15300252 {q_3}^2 \right]}{640640\sqrt[3] ar}+\frac{{q_3}^2+{q_4}^2}{r^2}+\frac{3a}{r^3}\nonumber\\
&\, -\frac{a^\frac{2}{3} \left( \sqrt{3}\pi \left[\sqrt{15300252} {q_4}^2-3326427 a^\frac{2}{3}+15300252 {q_3}^2 \right]+1281280 \sqrt[3] aM
+256256 a^\frac{4}{3}\Lambda \right) }{256256r^4}+ \frac{2a({q_3}^2+{q_4}^2)}{r^5} \nonumber\\
&\, +\frac{231a^2}{40r^6} - \frac{3a^\frac{5}{3} \left( \sqrt{3}
\pi \left[45900756 {q_3}^2+\sqrt{45900756} {q_4}^2-9979281 a^\frac{2}{3} \right] +1281280 a^\frac{4}{3}\Lambda+3843840 \sqrt[3] aM\right) }{1281280r^7}\nonumber\\
&\, +\frac{122a^2({q_3}^2+{q_4}^2)}{35r^8}+\frac{375a^3}{44r^9}\,,
\end{align}
where
\begin{align}
\Lambda=&\, \frac{10010 c_8 a^\frac{4}{3}-\sqrt{15300252} \sqrt{3}\pi {q_4}^2-15300252 \sqrt{3}
\pi {q_3}^2+3326427 c^\frac{2}{3}\sqrt{3}\pi }{2562560a^\frac{4}{3}}\,, \quad
q_3=c_2\, , \quad q_4=\sqrt{52998400}c_3\,,\nonumber\\
M=&\, -\frac{15300252 \sqrt{3}\pi {q_3}^2 +2562560 a^\frac{4}{3}\Lambda+160160 c^\frac{4}{3}c_7-3326427 c^\frac{2}{3}
\sqrt{3}\pi +\sqrt{15300252} \sqrt{3}\pi {q_4}^2}{1281280\sqrt[3] a}\, .
\end{align}
Eq.~(\ref{mpabB}) tells that the space-time asymptotically approaches to the AdS/dS space-time.
Compared with the metric given in general relativity, extra terms appear including the dimensional parameter $a$ in the metric
due to the contributions of higher-order curvature terms of $F(R)$ gravity.
We find that when the charges $q_3$ and $q_4$ vanish and the dimensional parameter $a$ vanishes,
the space-time reduces to the Schwarzschild one \cite{Misner:1973prb}.
In spite that we recover the Schwarzschild space-time in the limit of $q_3$, $q_4$, $a \to 0$,
we cannot recover the Reissner-Nordstr{\" o}m in any limit because the dimensional parameter $a$ cannot vanish
when there are non-vanishing charges $q_3$ and $q_4$, that is,
when the charges $q_3$ and $q_4$ do not vanish, the parameter $a$ is not allowed to vanish
as the second term of $h_1$ in Eq.~(\ref{mpabB}) shows.
This result is consistent with what has been obtained in \cite{Nashed:2021lzq}.

By using Eq.~(\ref{mpabB}), we calculate the invariants as follows,
\begin{align}
\label{inv}
R_{\mu \nu \rho \sigma} R^{\mu \nu \rho \sigma}=&\, 24 \Lambda^2-\frac{24 \Lambda \sqrt{3}\pi\left[\frac{347733}{29120}
\left( {q_4}^2+{q_3}^2 \right)- \frac{255879}{98560} a^\frac{2}{3}\right]}{\sqrt[3] ar^3} \nonumber \\
&\, -\frac{8 \left( 7 a^\frac{4}{3}\Lambda+\sqrt{3}\pi \left[{\frac{85293}{98560}} a^\frac{2}{3}-\frac{347733}{29120} \left( {q_4}^2+{q_3}^2 \right) \right] \right)}{\sqrt[3] ar^5}
\cdots\,, \nonumber\\
R_{\mu \nu } R^{\mu \nu }=&\, 36 \Lambda^2-\frac{36 \pi \sqrt{3}\Lambda \left( \frac{347733}{29120} \left( {q_4}^2+{q_3}^2 \right)
 - \frac{255879}{98560} a^\frac{2}{3} \right)}{\sqrt[3] ar^3} \nonumber \\
&\, -\frac{36 \left(\frac{7}{3} a^\frac{4}{3}\Lambda +\frac{85293}{98560} a^\frac{2}{3}\sqrt{3}\pi
 - \frac{115911 \pi \sqrt{3} \left( {q_4}^2+{q_3}^2 \right)}{29120} \right)}{\sqrt[3] ar^5} \cdots\,, \nonumber\\
R=&\, -12 \Lambda+\frac{767637 \pi\sqrt{3} \left[ \frac{583}{507} \left( {q_4}^2+{q_3}^2 \right)
 -\frac{1}{4} a^\frac{2}{3} \right]}{12320 \sqrt[3] ar^3} \nonumber \\
&\, +{\frac{767637 \left(\frac{1}{12} a^\frac{2}{3}\sqrt{3} \pi + \frac{172480}{767637}\Lambda a^\frac{4}{3}- \frac{583}{1521}
\sqrt{3} \left( {q_4}^2+{q_3}^2 \right) \pi \right)}{12320\sqrt[3] a\Lambda r^5}} \cdots\,.
\end{align}
{
Eq.~(\ref{inv}) shows that the dimensional parameter $a$ should not vanish, again, so that the invariants in (\ref{inv}) could be finite.
We stress that the parameter $a$ is the source of deviation from general relativity and we may compare the behaviors of the invariants in (\ref{inv})
with those of the invariants in general relativity in Eq.~(\ref{R1}).

Although the expressions in (\ref{inv}) are valid when $r$ is large,
by using the exact expressions (\ref{fr1}) and (\ref{frB1}) of the solution, we can find the behaviors of these scalar invariants when $r\sim 0$.
We repeat the above calculations for small $r$.
When $r$ is small, $h$ and $h_1$ behave as in the following form:
\begin{align}
\label{fr11}
h \approx \frac{a_1}{r^{24}}+\frac{a_2}{r^{23}}+\frac{a_3}{r^{21}}\,, \quad
h_1 \approx a_4+a_5r+a_6r^3+a_7r^4\,,
\end{align}
where 
\[
a_1=-\frac{a^8}{46592}\,,\quad   a_2=\frac{a^7 \left( {c_2}^2+52998400{c_7}^2 \right)}{20800}\,,\quad  a_3=\frac{197 a^7}{453376}\,,\quad  a_4=-\frac{1}{182}\,, 
\]
\[
a_5=\frac{4 \left( {c_2}^2+52998400{c_7}^2 \right)}{325a}\,,\quad  a_6=\frac{537}{23023a}\,, \quad  \textrm{ and} \quad  a_7=-\frac{266 \left( {c_2}^2+52998400 {c_7}^2 \right)}{3575a^2}\,.
\]
By using Eq. \eqref{fr11} to calculate the invariants,  we obtain
\begin{align}
\label{inv1}
R_{\mu \nu \rho \sigma} R^{\mu \nu \rho \sigma} \approx&\, \frac{58113}{8281 r^4}+\frac{a_8}{r^3}+\frac{a_9}{r^2}\,,\nonumber\\
R_{\mu \nu} R^{\mu \nu} \approx&\, \frac{50121}{16562r^4}+\frac{a_{10}}{r^3}+\frac{a_{11}}{r^2}\,,\nonumber\\
R \approx&\, \frac{45}{13r^2}+\frac{a_{12}}{r}\,,
\end{align}
where  
\[
a_8=-\frac{50256 \left( {c_2}^2+52998400 {c_7}^2 \right)}{4225a}\,, \quad  a_9=\frac{247168 \left( {c_2}^2+52998400 {c_7}^2 \right)^2}{21125a^2}\,, \quad 
a_{10}=-\frac{133776 \left( {c_2}^2+52998400 {c_7}^2 \right)^2}{29575a}\,,
\]
\[
a_{11}=\frac{523936 \left( {c_2}^2+52998400 {c_7}^2 \right)^2}{105625a^2}\,, \quad a_{12}=-\frac{72 \left( {c_2}^2+52998400 {c_7}^2 \right)}{25a}\,.
\]
Therefore all these scalar invariants have a true singularity at $r=0$ and the leading terms of the invariants
$\left(R_{\mu \nu \rho \sigma} R^{\mu \nu \rho \sigma},R_{\mu \nu} R^{\mu \nu},R\right)$ are given by
$\left( \mathcal{O}\left(\frac{1}{r^3}\right), \mathcal{O}\left(\frac{1}{r^3}\right), \mathcal{O}\left(\frac{1}{r^3}\right) \right)$, which are
different from the behaviors in the charged black hole in general relativity, where the leading term of the Kretschmann scalar
$R_{\mu \nu \rho \sigma} R^{\mu \nu \rho \sigma}$ is $\mathcal{O}\left( \frac{1}{r^6}\right)$, and
the other invariants behave as $R_{\mu \nu} R^{\mu \nu}=\mathcal{O}\left(\frac{1}{r^8}\right)$
and $R=\mbox{const.}$
Therefore, the singularities of the Kretschmann and the Ricci tensor squared are softer than those in
the charged black hole of general relativity.}

Because we are interested in the black hole that deviated from general relativity, in the next section, we will study the thermodynamics of the black hole (\ref{mpabB})

\section{Thermodynamics of the black hole in (\ref{mpabB})}\label{S4}


In this section, we will study the thermodynamical properties of the black hole given in (\ref{mpabB}).
For this purpose, we remind you of the basic definitions of thermodynamical quantities.
First, we depict the behavior of the metric potential $h(r)$ in Figure~\ref{Fig:1}~\subref{fig:met}.
 From Figure~\ref{Fig:1}~\subref{fig:met}, we find that there are two horizons, the inner Cauchy horizon $r=r_1$ and the outer horizon $r=r_2$,
when the parameter $a>-1.1$, these horizons coincides when the dimensional parameter $a=-1.1$
(or more exactly $a=-1.1\kappa$ although we have chosen the unit where $\kappa=1$), i.e.,
we have only one horizon which is called the degenerate horizon, $r_d$, and when $a<-1.1$, there appears the naked singularity.
Finally, we stress that although $h(r)=-g_{tt}$ does not equal to $h_1(r)=\frac{1}{g_{rr}}$, they have the same Killing and event horizons,
that is, when $h(r)=0$, we find $h_1(r)=0$ and vice versa.

The Hawking temperature $T$ is defined as \cite{Sheykhi:2012zz,Sheykhi:2010zz,Hendi:2010gq,Sheykhi:2009pf,Wang:2018xhw,Zakria:2018gsf},
\begin{align}
\label{temp}
T= \frac{h'\left(r_2\right)}{4\pi}\,.
\end{align}
Moreover, the Hawking entropy of the outer horizon is given by
\begin{align}
\label{ent}
S\left(r_2\right) =\frac{1}{4} A \left(r_2 \right) F_R\,.
\end{align}
Here $A\left(r_2\right)$ is the area of the outer horizon.

The local energy $E$ has the following form \cite{Cognola:2011nj,Sheykhi:2012zz,Sheykhi:2010zz,Hendi:2010gq,Sheykhi:2009pf,Zheng:2018fyn},
\begin{align}
\label{en}
E \left( r_2 \right)=\frac{1}{4}\int\left[2F_R\left(r_2\right)+{r_2}^2\left\{F\left(R \left(r_2\right)\right)-R \left(r_2\right)F_R \left(r_2\right) \right\}\right]dr_2\, .
\end{align}
Using the above thermodynamical quantities, we define the Gibbs free energy as follows \cite{Zheng:2018fyn, Kim:2012cma},
\begin{align}
\label{enr}
G\left(r_2\right)=E\left(r_2\right)-T\left(r_2\right)S\left(r_2\right)\, .
\end{align}

\subsection{The thermodynamics of the black hole solution (\ref{mpabB})}

\begin{figure}
\centering
\subfigure[~The metric potential of black hole (\ref{mpabB})]{\label{fig:met}\includegraphics[scale=0.3]{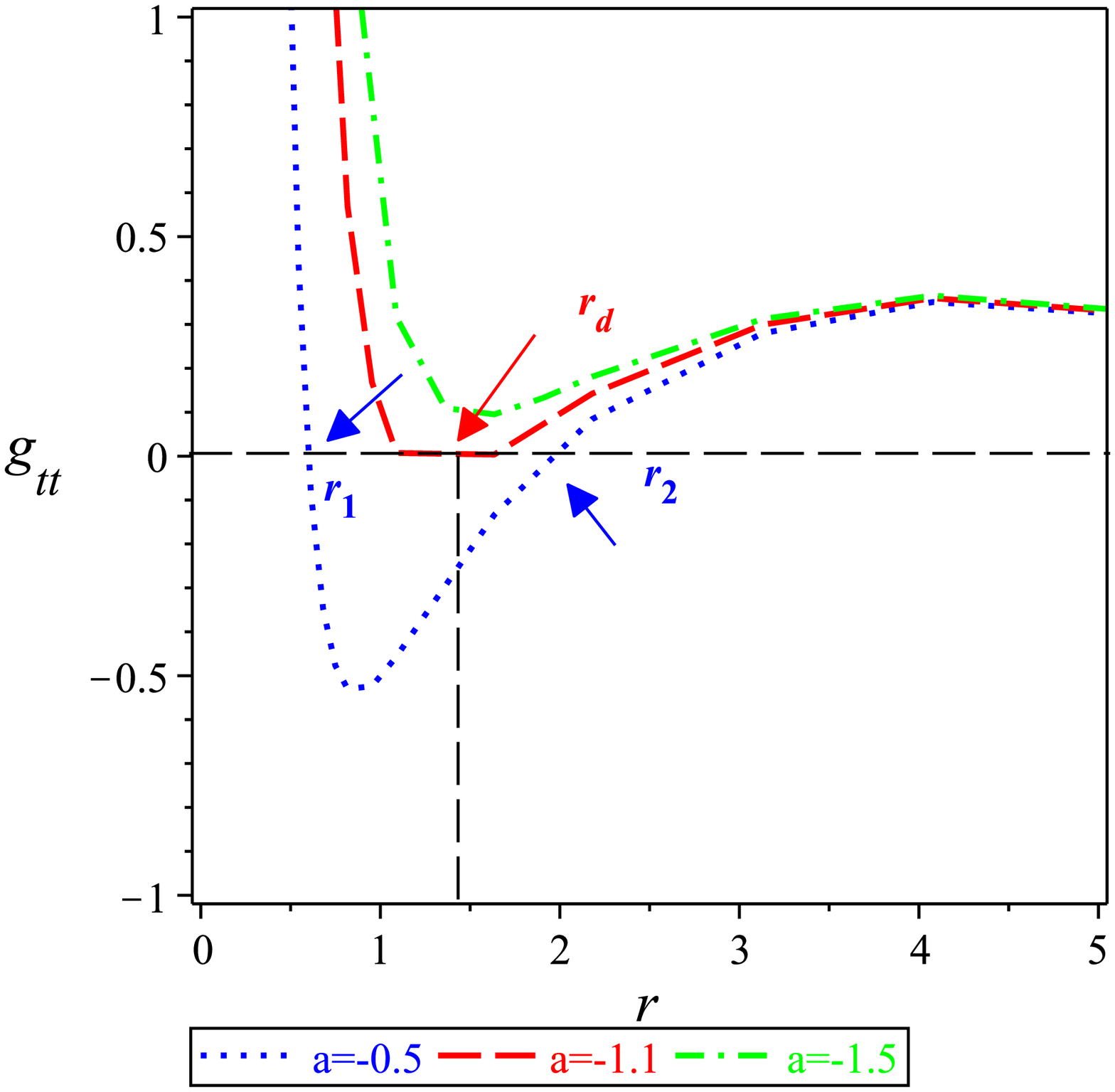}}
\subfigure[~The Hawking temperature of the black hole (\ref{mpabB}) when $a=-0.5$]{\label{fig:temp}\includegraphics[scale=0.3]{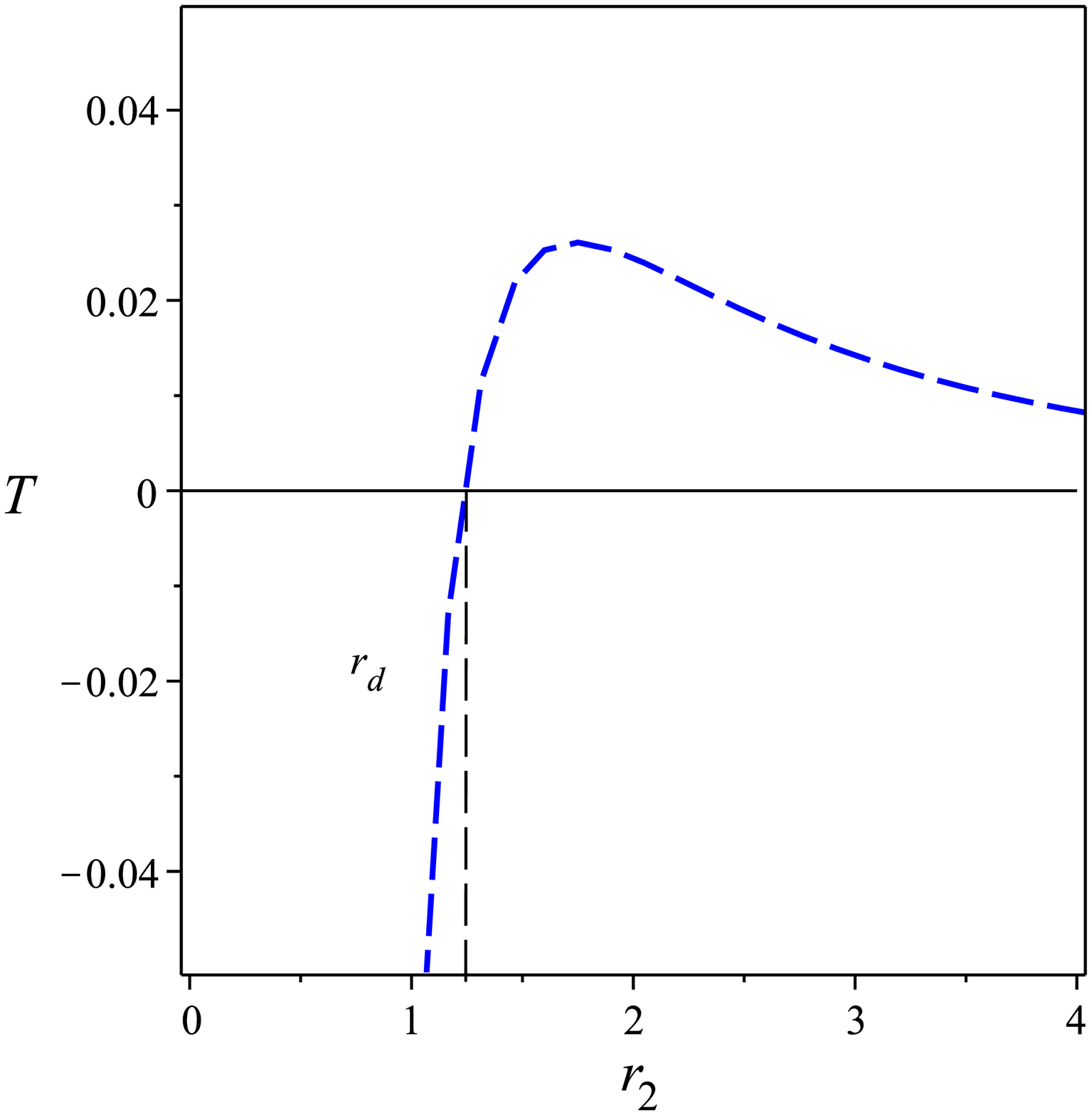}}
\subfigure[~The Hawking entropy of black hole (\ref{mpabB}) when $a=-0.5$]{\label{fig:ent11}\includegraphics[scale=0.3]{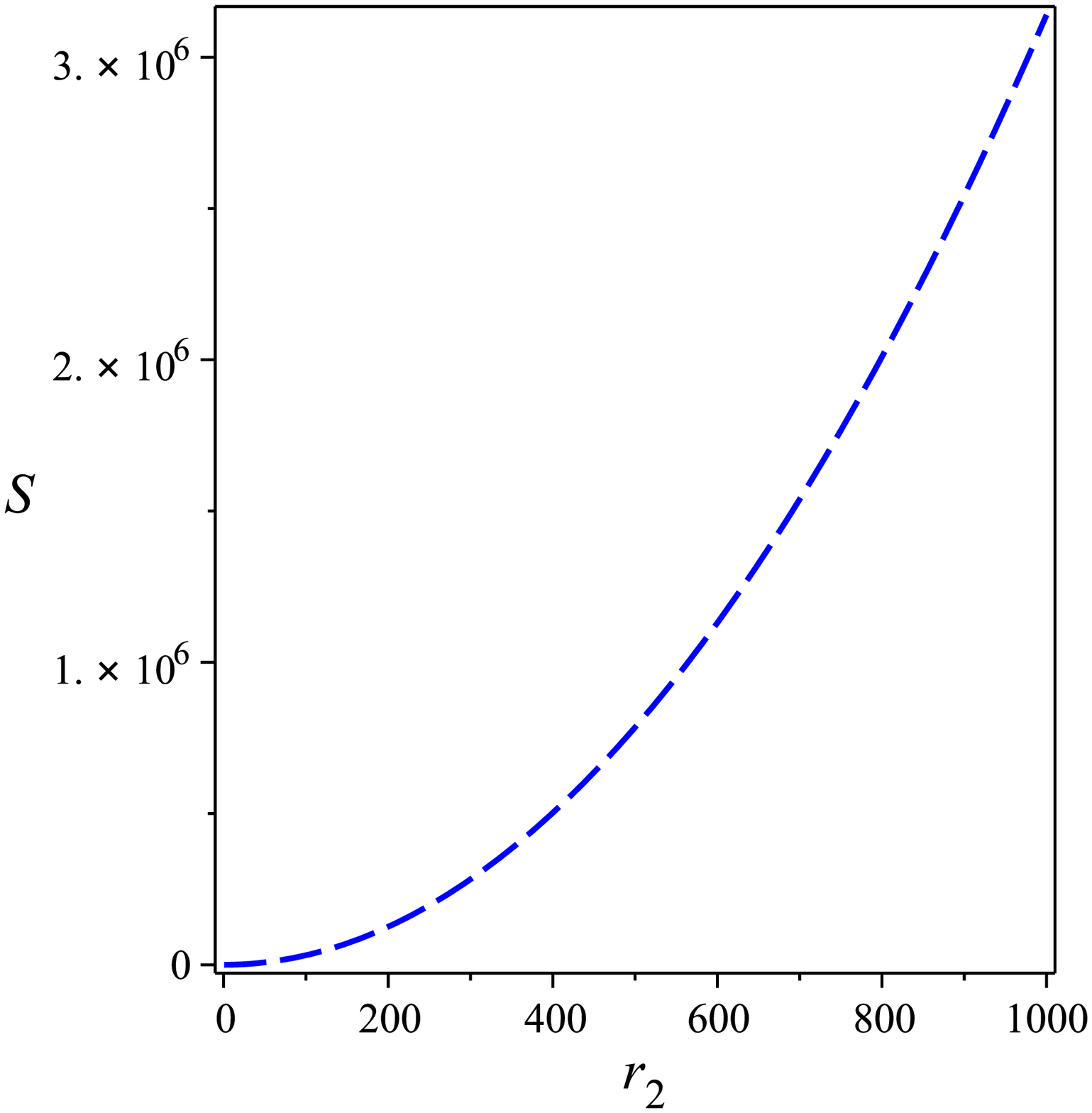}}
\subfigure[~The quasi-local energy of black hole (\ref{mpabB}) when $a=-0.5$]{\label{fig:enrg11}\includegraphics[scale=0.3]{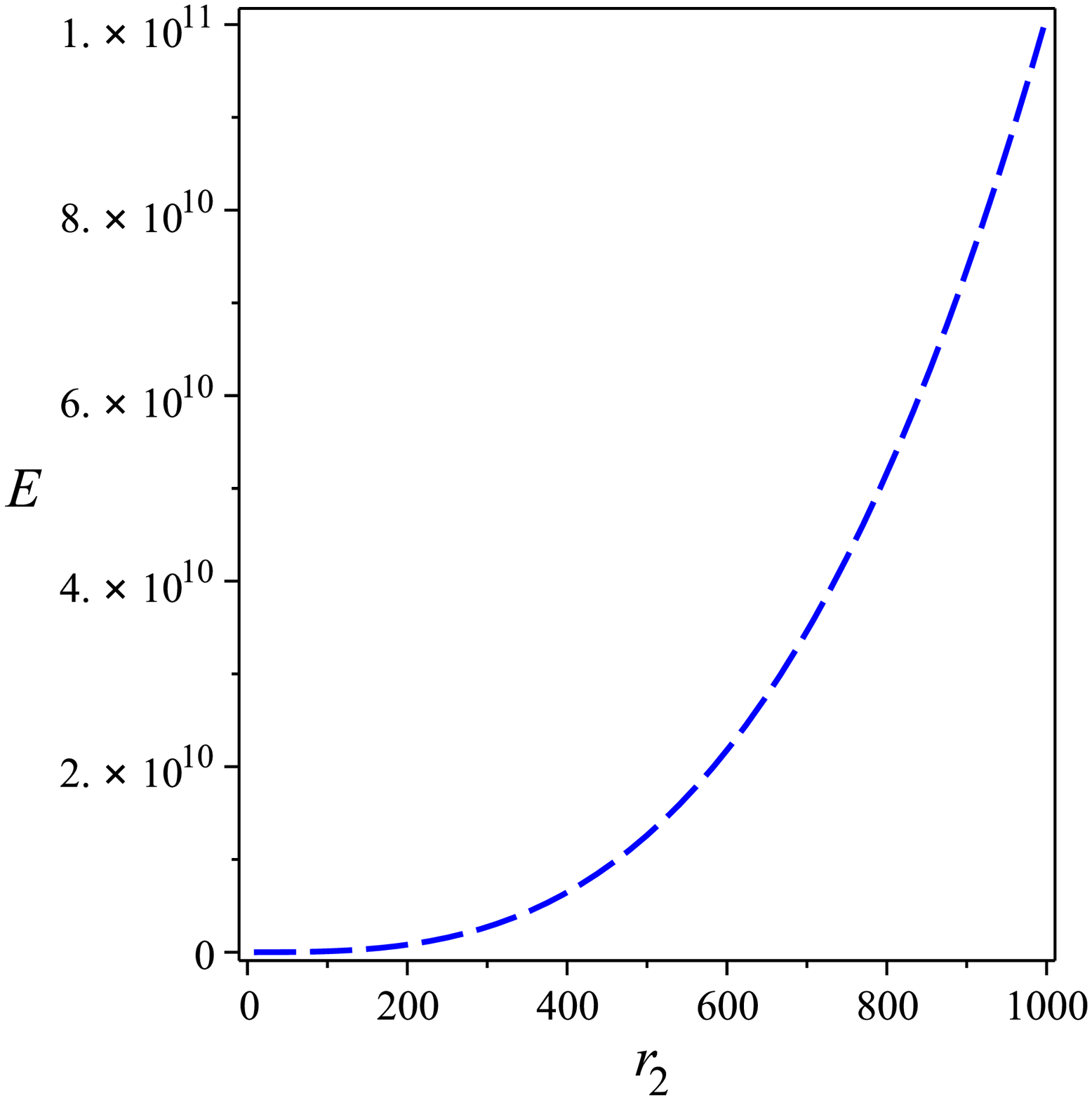}}
\subfigure[~The Gibbs free energy of black hole (\ref{mpabB}) when $a=-0.5$]{\label{fig:gib11}\includegraphics[scale=0.3]{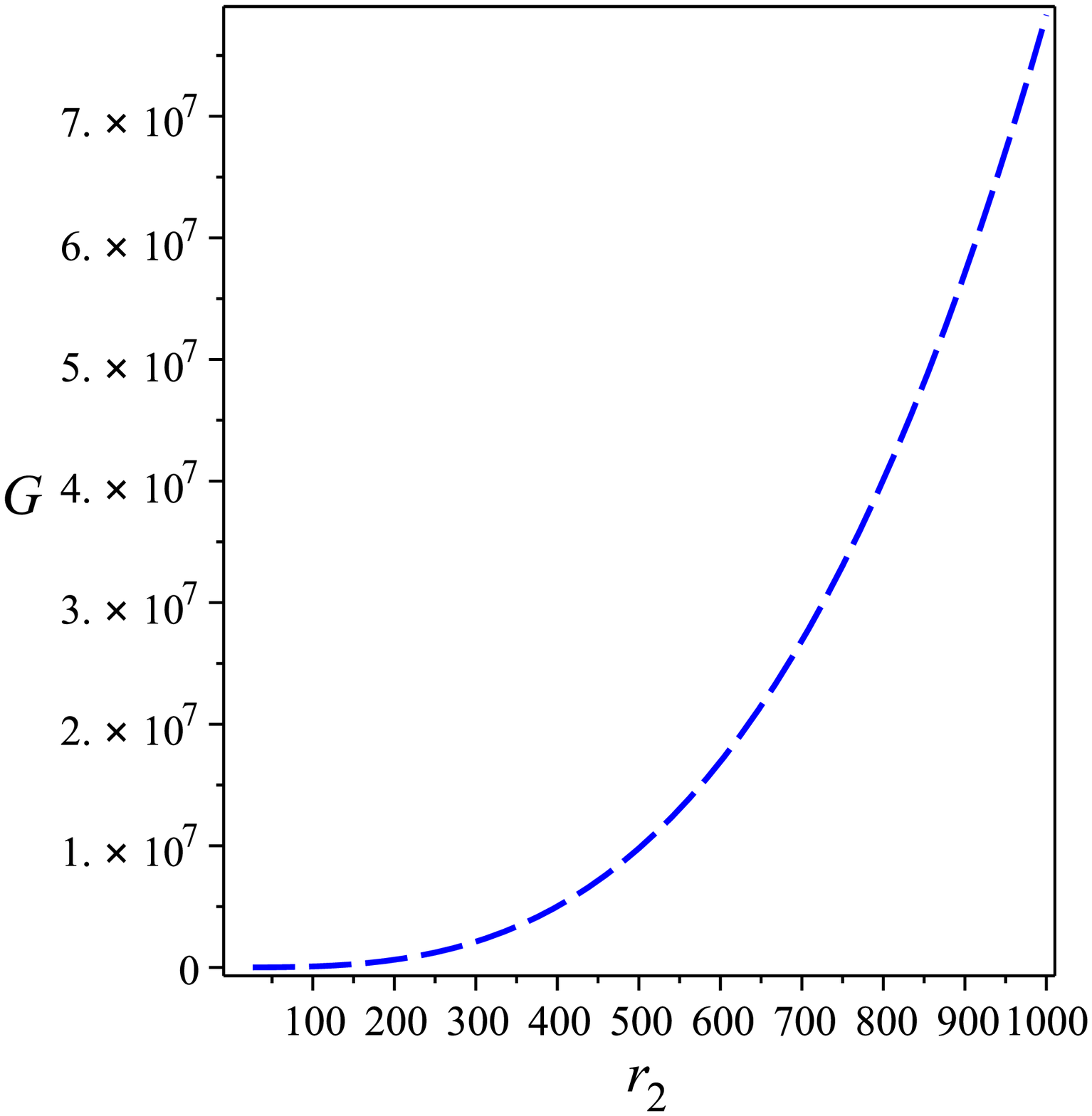}}
\caption[figtopcap]{\small{{Plot of the physical thermodynamics of the black hole solution (\ref{mpabB});
Figure~\subref{fig:met} gives the horizons given by Eq.~(\ref{mpabB}) of the metric potential $h$;
Figure~\subref{fig:temp} the Hawking temperature; Figure~\subref{fig:ent11} the Hawking entropy;
Figure~\subref{fig:enrg11} the quasi-local energy; Figure~\subref{fig:gib11} the Gibbs free energy.Here we take $M=1$, $\Lambda=0.01$, and $q_3=q_4=10^{-5}$.}}}
\label{Fig:1}
\end{figure}

The black hole solution given by Eq.~(\ref{mpabB}) is characterized by its mass $M$, the dimensional parameter $a$, and the electric and magnetic charges $q_3$ and $q_4$.
We stress that the dimensional parameter $a$ cannot vanish and therefore we cannot recover the Reissner-Nordstr{\" o}m solution in any limit.
However, when we set the electric and magnetic field equal to zero and then set $a=0$, we can recover the Schwarzschild solution which corresponds to general relativity.
To find the horizon radii of the black hole (\ref{mpabB}), we solve the equation $h(r)=0$ but we keep the order of $\frac{1}{r}$ up to $\cal{O}\left(\frac{1}{r^3}\right)$
for an approximation.
Under the approximation, the equation $h(r)=0$ has five roots, two of them are real and the other three are pure imaginary.
It is difficult to find the analytic forms of these roots easy but the asymptotic forms of the real roots are drawn numerically
in Figure~\ref{Fig:1}~\subref{fig:met}. From Figure~\ref{Fig:1}~\subref{fig:met}, we find that there are surely two horizons where $h(r)=0$.
Using Eq.~(\ref{temp}), we calculate the Hawking temperature and obtain
\begin{align}
\label{T1}
T\left(r_2\right)=\frac{2\Lambda {r_2}^5+2M{r_2}^2-2 \left( {q_3}^2+{q_4}^2 \right)r_2+3a}{4\pi {r_2}^4}\,.
\end{align}
The behavior of the temperature given by Eq.~(\ref{T1}) is depicted in Figure~\ref{Fig:1}~\subref{fig:temp}, which shows that
$T_2\equiv T\left(r_2\right) >0$ as far as $r_2>r_d$.
Figure \ref{Fig:1}~\subref{fig:temp} indicates that temperature $T$ vanishes when $r_2 = r_d$.
Moreover, when $r_2 < r_d$, the temperature $T$ becomes negative, and an ultracold black hole is constructed but because there appears a naked singularity,
the case $r_2 < r_d$ could be prohibited by cosmic censorship, that is, the space-time with the naked singularity could not be generated by the collapse of the matters.
On the other hand, Davies \cite{Davies:1977bgr} claimed that there is no concrete reason from the view of thermodynamical effects
to stop black hole temperature to take negative values or to create a naked singularity.
Figure \ref{Fig:1}~\subref{fig:temp} seems to support Davies' argument at $r_2 < r_d$ region.

By using Eq.~(\ref{ent}), we obtain the entropy of black hole (\ref{mpabB}) as follows,
\begin{align}
\label{S1}
S\left(r_2\right)= \frac{\pi \left( {r_2}^3+a \right)}{r_2}\,.
\end{align}
The behavior of the entropy given by Eq.~(\ref{S1}) is depicted in Figure~\ref{Fig:1}~\subref{fig:ent11} which indicates that $S\left(r_2\right)$ is surely positive and
an increasing function of $r_2$.

 From Eq.~(\ref{en}), we find that the quasi-local energy takes the following form,
\begin{align}
\label{E1}
E\left(r_2\right)=&\, \frac{1}{38438400\Lambda {r_2}^5} \left\{45900756 \sqrt{3} \left({q_4}^2+{q_3}^2 \right) \pi
\left(5\Lambda {r_2}^2-1 \right) a^\frac{2}{3}-9979281 \sqrt{3}\pi \left( 5\Lambda\,{r_2}^2-1 \right) a^\frac{4}{3} \right. \nonumber\\
&\, \left. +32032 \Lambda \left(60 {r_2}^6-30 a {r_2}^3+10 {r_2}^8c_8+120 {r_2}^8\Lambda+360\Lambda a\ln \left(r_2\right) {r_2}^5+84 a^2 \right) \right\} \,.
\end{align}
The behaviors of the quasi-local energies are shown in Figure~\ref{Fig:1}~\subref{fig:enrg11} which also shows that $E\left(r_2\right)$ is also positive
and an increasing function of $r_2$.
Finally, we use Eqs.~(\ref{T1}), (\ref{S1}), and (\ref{E1}) in Eq.~(\ref{enr}) to calculate the Gibbs free energies and obtain
\begin{align}
G\left(r_2\right)=&\, \frac{1}{38438400\Lambda r^5} \left\{45900756 \left( 5\Lambda r^2-1 \right) \left( {q_3}^2+{q_4}^2 \right) \pi \sqrt{3} a^\frac{2}{3}
 -9979281 \left( 5 \Lambda r^2-1 \right) \pi \sqrt{3} a^\frac{4}{3} \right. \nonumber\\
& \left. +320320 \Lambda \left(60 r^6 +360 \Lambda a\ln ( r) r^5-30 ar^3+ \left( 10 c_8+120\,\Lambda \right) r^8+84 a^2 \right)\right\} \nonumber \\
&\, -\frac{\left( 2 \Lambda r^5+2 Mr^2-2 r \left( {q_3}^2+{q_4}^2 \right) +3 a \right) \left( r^3+a \right) }{4r^5}\,.
\end{align}
The behavior of this free energy is depicted in Figure~\ref{Fig:1}~\subref{fig:gib11} which shows that $G_2$  is positive
and an increasing function of $r_2$, again.
\begin{figure}
\centering
\subfigure[~The metric potential of black hole (\ref{mpab1})]{\label{fig:metm}\includegraphics[scale=0.3]{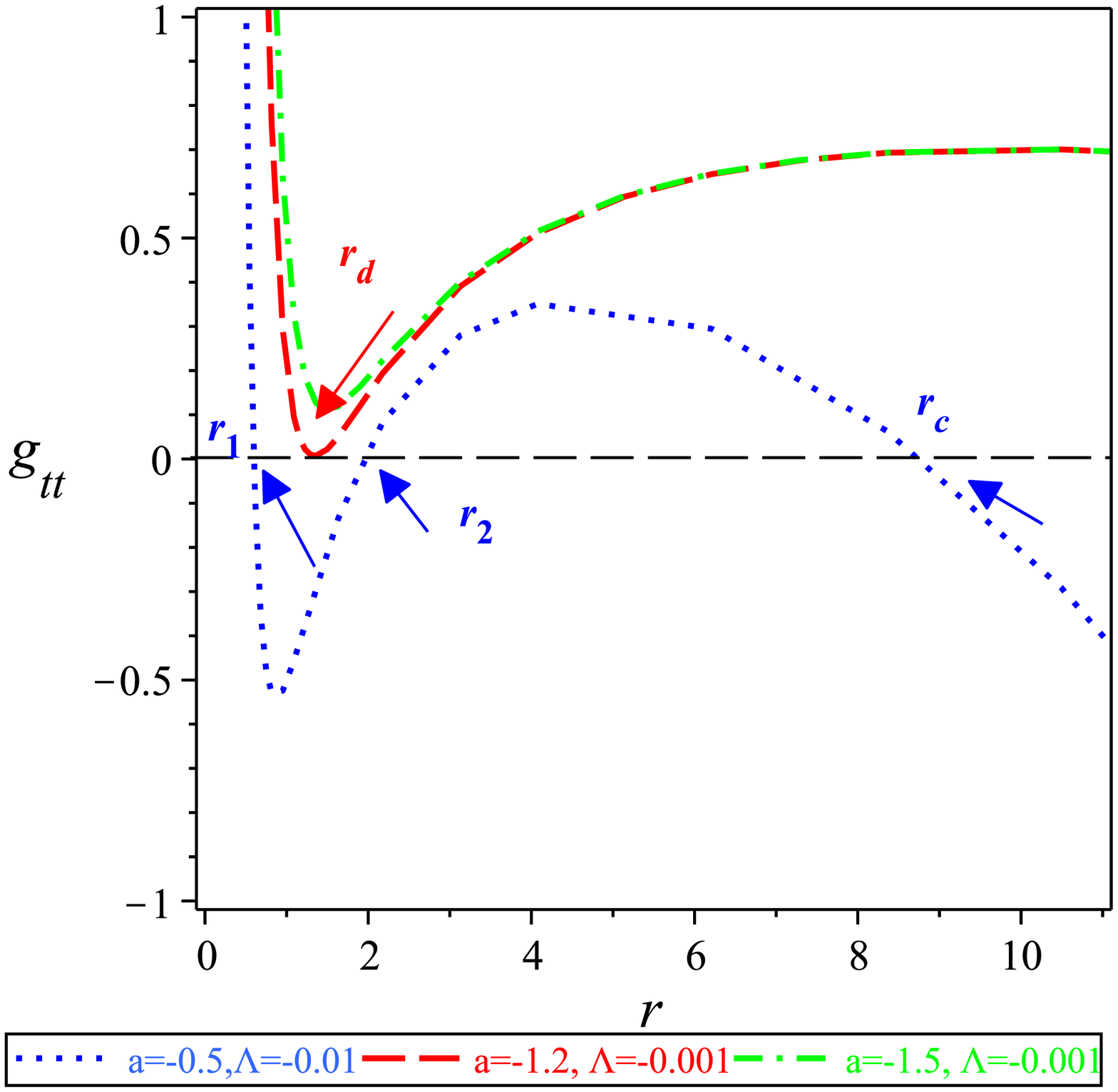}}
\subfigure[~The Hawking temperature of the black hole (\ref{mpab1}) when $a=-0.5$]{\label{fig:tempm}\includegraphics[scale=0.3]{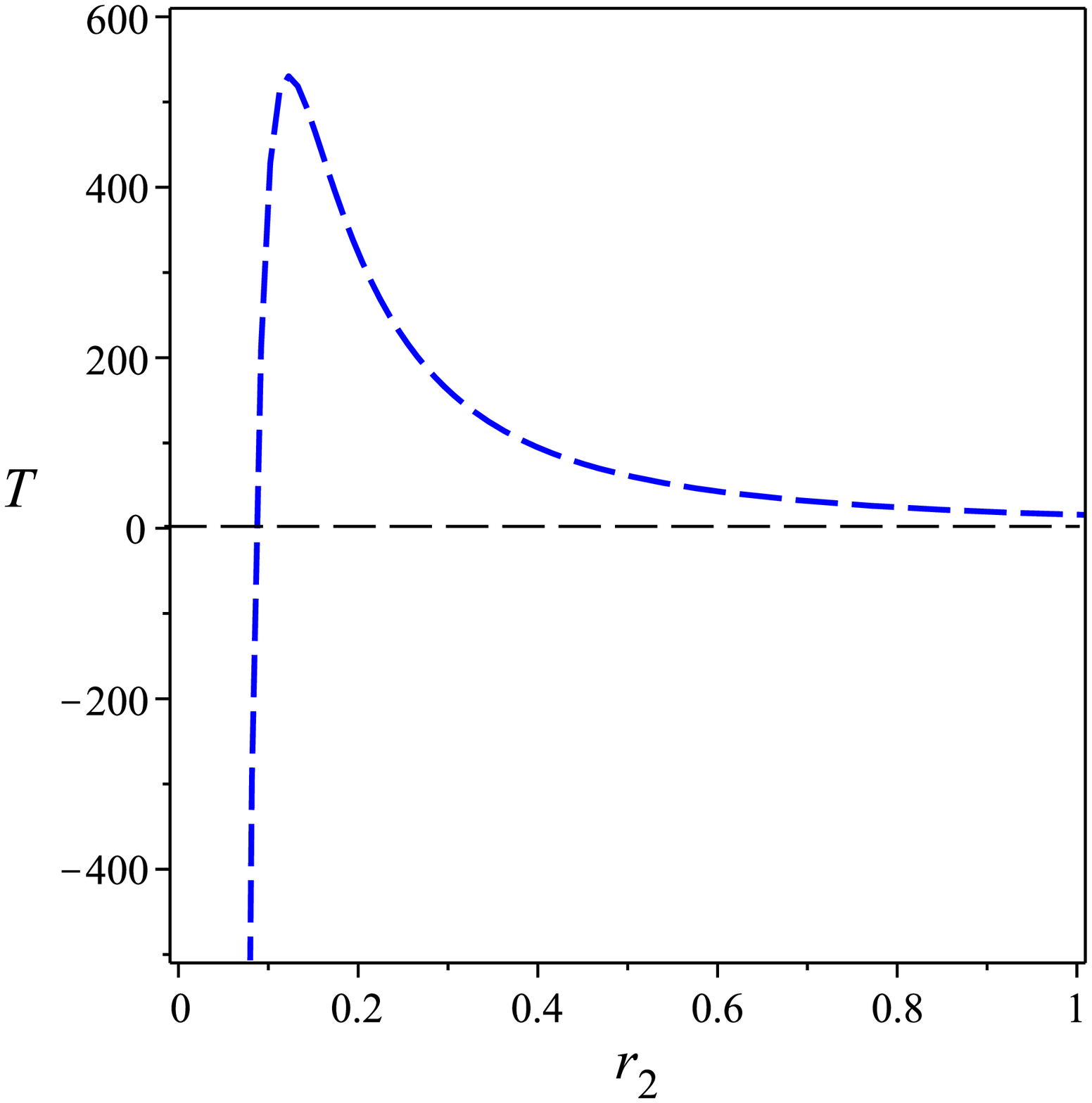}}
\subfigure[~The quasi-local energy of the black hole (\ref{mpab1})when $a=-0.5$ ]{\label{fig:enrm}\includegraphics[scale=0.3]{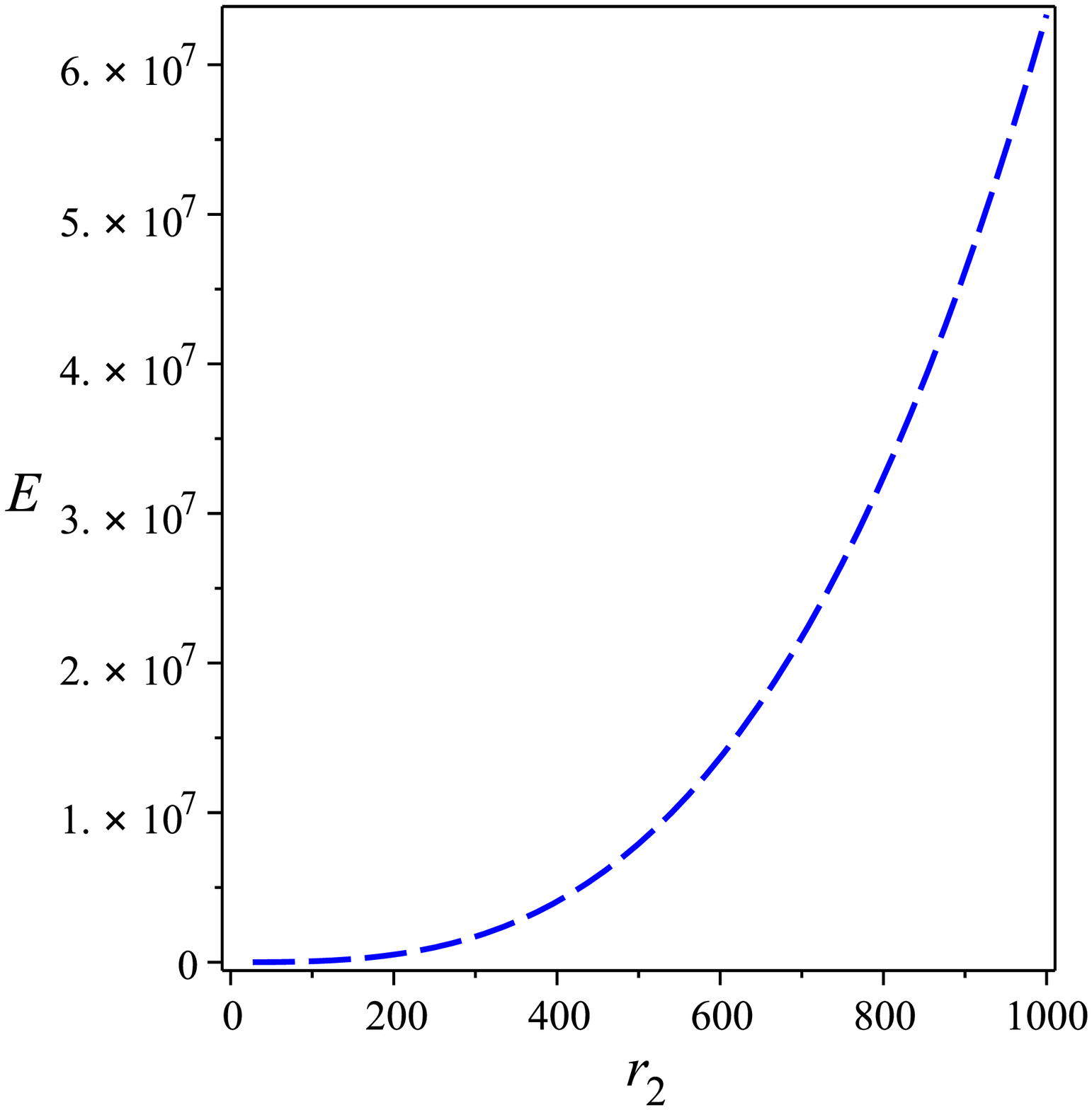}}
\subfigure[~The Gibbs free energy of the black hole (\ref{mpab1}) when $a=-0.5$]{\label{fig:gibm}\includegraphics[scale=0.3]{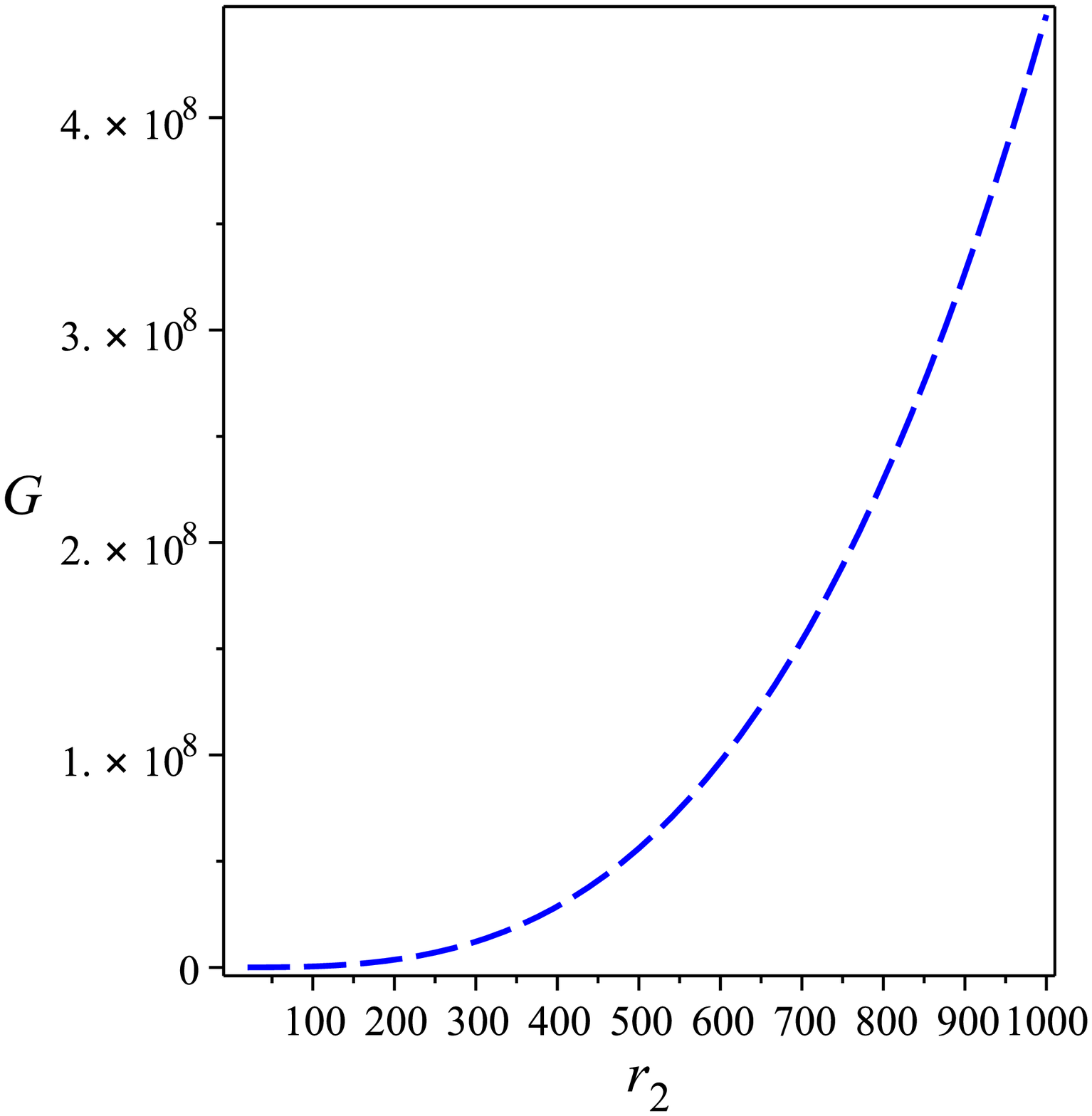}}
\caption[figtopcap]{\small{{Plot of the physical thermodynamics of the black hole solution (\ref{mpabB});
Figure~\subref{fig:met} gives the horizons given by Eq.~(\ref{mpabB}) of the metric potential $h$;
Figure~\subref{fig:temp} the Hawking temperature;
Figure~\subref{fig:ent11} the Hawking entropy;
Figure~\subref{fig:enrg11} the quasi-local energy;
Figure~\subref{fig:gib11} the Gibbs free energy.Here we take $M=1$, and $q_3=q_4=10^{-5}$.}}}
\label{Fig:2}
\end{figure}

\subsection{Thermodynamics of multi-horizon black hole (\ref{mpabB}) with negative value of the cosmological constant}

In the previous subsection, we studied a black hole solution with two horizons where the cosmological constant is positive.
In this subsection, we study the same black hole solution but with a negative cosmological constant $\Lambda$ which generates a cosmological horizon and
we obtain the black hole solution with three horizons.
The solution is also characterized by the mass $M$, the parameter $a$, and the electric and magnetic charges $q_3$ and $q_4$.
When $\Lambda$ is negative, the metric components $h(r)=-g_{tt}$ and $h_1(r)=\frac{1}{g_{rr}}$ take the following forms,
\begin{align}
\label{mpab1}
h(r)\approx&\, -\Lambda r^2+ 1-\frac{2M}r+\frac{{q_3}^2+{q_4}^2}{r^2}-\frac a{r^3}-\frac{2a \left( {q_4}^2+q_3^2 \right) }{r^5}\,,\nonumber\\
h_1(r)\approx&\, -\Lambda r^2+ 1-\frac{1281280 \sqrt[3] aM-\sqrt{3}\pi \left[3326427 a^\frac{2}{3}
 -\sqrt{15300252}{q_4}^2-15300252 {q_3}^2 \right]}{640640\sqrt[3] ar}+{\frac{{q_3}^2+{q_4}^2}{r^2}}+\frac{3a}{r^3}\nonumber\\
&\, -\frac{a^\frac{2}{3} \left( \sqrt{3}\pi \left[\sqrt{15300252}{q_4}^2-3326427 a^\frac{2}{3}+15300252 {q_3}^2\right]+1281280 \sqrt[3] aM-256256 a^\frac{4}{3}\Lambda \right) }{256256r^4} \nonumber \\
&\, +{\frac{2a({q_3}^2+{q_4}^2)}{r^5}}\,.
\end{align}
When the electric charge $q_3$, the magnetic charge $q_4$, and the parameter $a$ vanish,
the space-time reduces to the Schwarzschild AdS space-time in Einstein's general relativity.
The behavior of $h(r)$ in the black hole geometry (\ref{mpab1}) is drawn in Figure~\ref{Fig:2}~\subref{fig:metm}.
From Figure~\ref{Fig:2}~\subref{fig:metm}, we find that there appear three horizons in general, where $h(r)=0$ in Eq.~(\ref{mpab1}) \cite{Wang:2018xhw}.
Although there are five roots in the equation $h(r)=0$, three of them are real but the others are imaginary.
The expressions of these real roots are lengthy, however, their behaviors are numerically drawn in Figure~\ref{Fig:2}~\subref{fig:metm}.
From the Figure, we find that the degenerate horizon for the metric potential $h(r)$ given by Eq.~(\ref{mpab1}) appears for specific values
for $(a,M,r)\equiv(-1.2,1,1.333717999)$, which corresponds to the Nariai black hole.

Using Eq.~(\ref{temp}), we obtain the Hawking temperature of the black hole (\ref{mpab1}) as given by Eq.~(\ref{T1}), whose behavior
is shown in Figure~\ref{Fig:2}~\subref{fig:tempm}, which shows that $T\left(r_2\right)>0$ as long as $r_2>r_d$.
Figure~\ref{Fig:2}~\subref{fig:tempm} also shows that $T\left(r_2\right)=0$ at $r_2 = r_d$.
When $r_2 < r_d$, $T\left(r_2\right)<0$ and an ultracold black hole might be formed.

By using Eq.~(\ref{en}), we draw the quasi-local energy in Figure~\ref{Fig:2}~\subref{fig:enrm}.
The Figure tells that the quasi-local energy is positive and an increasing function of $r_2$.
Finally, using Eqs.~(\ref{T1}), (\ref{S1}), and (\ref{E1}) in Eq.~(\ref{enr}), we find the Gibbs free energies.
The behaviors of these free energies are shown in Figure~\ref{Fig:2}~\subref{fig:gibm}, which tells us the free energy
is positive and an increasing function of $r_2$, again.

\subsection{First law of thermodynamics in the black holes (\ref{mpabB}) and (\ref{mpab1})}

It could be interesting to examine if the first law for the black hole geometry (\ref{mpabB}) could be verified.
The first law should have the following form even in the $F(R)$ gravity \cite{Zheng:2018fyn},
\begin{align}
dE=TdS-PdV\, ,
\label{1st}
\end{align}
where $E$ is the quasi-local energy, $S$ is the Bekenstein-Hawking entropy, $T$ is the Hawking temperature, $P$ is the radial component
of the stress-energy tensor that serves as a thermodynamic pressure $P=\left. {T_r}^r \right|_{\pm}$,
and $V$ is the geometric volume.
In the framework of the $F(R)$ gravity, the pressure can be defined as \cite{Zheng:2018fyn}
\begin{align}
P=-\frac{1}{8\pi}\left\{\frac{F_R}{{r_{\pm}}^2}+\frac{1}{2} \left( F - RF_R \right)\right\}+\frac{1}{4}\left(\frac{2F_R}{r_{\pm}}+F_R'\right)T\,.
\label{1stB}
\end{align}
For the space-time (\ref{mpabB}), if we neglect $O\left(\frac{1}{r^6} \right)$, we obtain
\begin{align}
\label{1stl}
E=&\, \frac{1}{38438400\Lambda {r_2}^5} \left\{45900756 \sqrt{3} \left({q_4}^2+{q_3}^2 \right) \pi
\left(5\Lambda {r_2}^2-1 \right) a^\frac{2}{3}-9979281 \sqrt{3}\pi \left( 5\Lambda {r_2}^2-1 \right) a^\frac{4}{3} \right. \nonumber\\
&\, \left. +32032 \Lambda \left(60{r_2}^6-30a{r_2}^3+10{r_2}^8c_8+120{r_2}^8\Lambda+360\Lambda a {r_2}^5 \ln r_2 +84 a^2 \right) \right\}\,,\nonumber\\
S=&\, \frac{\pi({r_2}^3+a)}{r_2}\,, \quad T=\frac{2M{r_2}^2+2\Lambda {r_2}^5-2 \left({q_3}^2+{q_4}^4 \right)r_2+3a}{4 \pi {r_2}^4}, \nonumber\\
P=&\, - \frac{1}{10250240{r_2}^8\Lambda \sqrt[3] a\pi }
\left\{\Lambda\sqrt[3] a \left[ 8968960 \Lambda a{r_2}^5 - 7047040 a^2+5125120 \Lambda{r_2}^8-2562560 {r_2}^3 a \right] \right. \nonumber\\
&\, \left. -3326427 a^\frac{5}{3}\sqrt{3}\pi +15300252 \sqrt{3}\pi a \left[ {q_4}^2+{q_3}^2 \right]+1281280 {r_2}^6\Lambda\sqrt[3] a
+9979281 \sqrt{3}\pi \Lambda{r_2}^2 a^\frac{5}{3} \right]\nonumber\\
&\, -45900756 \sqrt{3}\pi \Lambda{r_2}^2a \left[{q_4}^2+{q_3}^2 +\Lambda\sqrt[3] a\left(640640 c_8 {r_2}^8-2562560 M{r_2}^5
+1281280 a M{r_2}^2 \right. \right. \nonumber \\
&\, \left. \left. +2562560 {r_2}^4 \left[{q_4}^2+{q_3}^2 \right]-1281280\, a r_2 \left[ {q_4}^2+{q_3}^2 \right]\right)\right\}\,.
\end{align}
By substituting the equations in (\ref{1stl}) into (\ref{1st}), we have verified the first law of thermodynamics for the black hole (\ref{mpabB}).
Furthermore, by repeating the same procedure for the black hole (\ref{mpab1}), we verify the first law of thermodynamics, again

\section{The stability of the black holes}\label{S616}

In order to check the stability of the above black hole solutions given by Eqs.~(\ref{mpabB}) and (\ref{mpab1}), we rewrite the action Eq.~(\ref{a2}) of the  $F(R)$ gravity
in the scalar-tensor form
\begin{align}
S_\mathrm{g}=\frac{1}{2\kappa^2}\int d^4x\sqrt{-g} \left[\psi R-V(\psi) \right]\, ,
\label{action2}
\end{align}
where $\psi$ is a scalar field coupled to the Ricci scalar $R$ and $V(\psi)$ is the potential (see \cite{Capozziello:2011et,DeFelice:2011ka} for details).
To discuss the perturbation, we assume that the background is given by the spherically symmetric metric as follows,
\begin{align}
ds^2=g_{\mu\nu}^{(0)} dx^\mu dx^\nu =-h(r) dt^2+\frac{dr^2}{h_1(r)}+ r^2 \sum_{a,b=1}^2 \Omega_{ab} dx^a dx^b\, , \quad
\sum_{a,b=1}^2 \Omega_{ab} dx^a dx^b \equiv d\theta^2 +\sin^2\theta d\phi^2 \, .
\end{align}
Here we denote the background metric by $g_{\mu\nu}^{(0)}$.
We check the stability of the black hole solutions, (\ref{mpabB}) and (\ref{mpab1}), by using the linear perturbations.
Moreover, we are going to investigate the value of the propagation speed of the parity-odd perturbation.
For the action (\ref{action2}), the background equations have the following forms,
\begin{align}
V =- \frac{4h_1 \psi'}{r}-\frac{2\psi h_1 h'}{hr}-\frac{\psi' h_1 h'}{h}+\frac{2\psi}{r^2} -\frac{2h_1 \psi}{r^2}\, , \quad
\psi'' =-\frac{\psi'h'_1}{2h_1}-\frac{\psi h'_1}{r h_1}+\frac{\psi'h'}{2h}+\frac{\psi h'}{rh}\,, \quad R =\frac{dV}{d\psi}\,,
\label{eq:Ueq}\\
\end{align}
where $'$ means the differentiation w.r.t. the radial coordinate, $r$.

\subsection{Regge-Wheeler-Zerilli formulation}

Following Regge, Wheeler \cite{Regge:1957td}, and Zerilli \cite{Zerilli:1970se}, we decompose
the perturbed metric by using the decomposition under the two-dimensional rotations.
This decomposition is familiar in the perturbations of the Schwarzschild black hole solution in general relativity.

We consider the perturbation of $g_{\mu\nu}$ around the background metric $g_{\mu\nu}^{(0)}$, as follows,
\begin{align}
\label{pertl}
g_{\mu\nu}= g_{\mu\nu}^{(0)}+l_{\mu\nu}\,,
\end{align}
Here $l_{\mu\nu}$ corresponds to the perturbation and we assume the perturbed quantities to be much smaller than the background,
i.e., $\left| g_{\mu\nu}^{(0)}\right|\gg \left|l_{\mu\nu}\right|$.
Under the two-dimension rotations, $l_{tt}, l_{tr}$, and $l_{rr}$ transform as scalar with spin 0 while $l_{ta}$
and $l_{ra}$ transform as vector with spin 1 and $l_{ab}$ transforms as a tensor with spin 2.
Any function $\Psi$ including $\theta$ and $\phi$ can be expanded by the spherical harmonics $Y_{lm}\left( \theta, \phi\right)$,
\begin{align}
\Psi \left(t,r,\theta,\phi \right)=\sum_{l=0, 1, 2, \cdots} \sum_{m=-l, -l+1, \cdots, l} \Psi_{lm}\left(t,r \right)Y_{lm} \left( \theta,\phi \right)\, .
\label{scalar-decomposition}
\end{align}
The spherical harmonics $Y_{lm}\left( \theta, \phi\right)$ satisfies the following equation,
\begin{align}
\Delta_{\theta, \varphi}Y_{lm} \left( \theta, \phi \right)=- l \left( l+1 \right)Y_{lm} \left( \theta, \varphi \right)\, .
\end{align}
Here $\Delta_{\theta, \varphi}$ is the Laplacian on the two-dimensional unit sphere whose metric is given by $\Omega_{ab}$.
Additionally, the vector $P_a$ can decompose into two parts, that is, the gradient part and the rotational part by using two scalar functions $\Psi_1$
and $\Psi_2$, as follows,
\begin{align}
P_a \left(t,r,\theta,\phi \right)=\nabla_a \Psi_1+ E_a^{\ b}\nabla_b \Psi_2\,,
\end{align}
where $\nabla_ a$ is the covariant derivative with respect to the metric $\Omega_{ab}$ and
\begin{align}
\label{Eab}
E_{ab}\equiv \epsilon_{ab} \sqrt{\det\Omega}\, .
\end{align}
Here $\epsilon_{ab}$ is a second-order skew-symmetric tensor with
$\epsilon_{12}=1$ and $\epsilon_{ab}=-\epsilon_{ba}$.

For $K_{ab}$, which is a symmetric tensor, can be decomposed as
\begin{align}
K_{ab}(t,r,\theta,\varphi)=\nabla_a\nabla_b \Psi_3+\Omega_{ab}\Psi_4
+\frac{1}{2}\left({E_a}^c\nabla_c\nabla_b\Psi_{5} + {E_b}^c \nabla_c \nabla_ a\Psi_{5}\right)\,.
\end{align}
Here $\Psi_3$, $\Psi_4$, and $\Psi_5$ are three scalar functions.
Because $K_{ab}$ has three independent components that describe by $K_{ab}$, we can decompose the tensor $K_{ab}$
to $\Psi_3$, $\Psi_4$, and $\Psi_5$.  to decompose $K_{ab}$.
The importance of this decomposition is that in the linearized forms of motion, parity odd-type and parity even-type perturbations are
completely decoupled.
Note $\Psi_1$, $\Psi_3$, $\Psi_4$ are parity-even and therefore they are real scalars but $\Psi_2$ and $\Psi_5$ are parity-odd and pseudo scalars.

In the next subsection, we will study the odd-type perturbations.

\subsection{Perturbations with respect to odd-modes}

In the Regge-Wheeler formalism, the odd-type metric perturbations have the following form,
\begin{align}
l_{tt}=&\, 0\, ,\quad l_{tr}=0\, , \quad l_{rr}=0\, ,\nonumber \\
l_{ta}=&\, \sum_{l , m}l_{\mathrm{0},l m}(t,r)E_{ab}\partial^{b}Y_{l m}(\theta,\varphi)\, ,\nonumber \\
l_{ra}=&\, \sum_{l , m}l_{\mathrm{1},l m}(t,r)E_{ab}\partial^{b}Y_{l m}(\theta,\varphi)\, ,\nonumber \\
l_{ab}=&\, \frac{1}{2}\sum_{l , m}l_{\mathrm{2},l m}(t,r)\left[E_ a^{~c}\nabla_ c\nabla_{b}Y_{l m}(\theta,\varphi)
+E_{b}^{~c}\nabla_ c\nabla_ aY_{l m}(\theta,\varphi)\right]\, .
\end{align}
By using the gauge transformation $x^{\mu}\to x^{\mu}+\xi^{\mu}$, one can choose some of the components in the metric perturbations to vanish.
We now use the following transformation of the odd-type perturbation,
\begin{align}
\xi_t = \xi_r=0\, , \quad \xi_ a = \sum_{l, m}\Lambda_{lm} \left(t,r \right) {E_ a}^b \nabla_b Y_{lm}\, ,
\end{align}
and we choose $\Lambda_{lm}$ so that $l_{\mathrm{2}, lm}$ vanishes, which is called the Regge-Wheeler gauge.
By using the gauge condition, the action of odd modes becomes \cite{Regge:1957td}
\begin{align}
S_\mathrm{odd} =&\, \frac{1}{2\kappa^2} \sum_{n,m}\int dt dr \mathcal{I}_\mathrm{odd} \nonumber \\
=&\, \frac{1}{4\kappa^2} \sum_{n,m}\int dt dr j^2 \left[\psi\sqrt{\frac{h_1}{h}}{\left(\dot{l}_1-l_0'\right)}^2
+\frac{4l_0{\dot{l}_1} \psi}r\sqrt{\frac{h_1}{h}}+ \frac{l_0^2}{r^2}\left[2\left(r\left\{\psi \sqrt{\frac{h_1}{h}}\right\}' +\psi\sqrt{\frac{h_1}{h}}\right)
\right. \right. \nonumber\\
& \left. \left. +\frac{\left(j^2-2\right)\psi}{\sqrt{hh_1}} \right]-\frac{\left( j^2-2 \right) \sqrt{hh_1} \psi l_1^2}{r^2}\right]\, ,
\label{odd-action}
\end{align}
where $j^2=l(l+1)$.
We should note that the action (\ref{odd-action}) does not include the derivative of $l_0$ concerning time, that is ${\dot l}_0$, and therefore
$l_0$ is not a dynamical degree of freedom.
We rewrite $\mathcal{I}_\mathrm{odd}$ in Eq.~(\ref{odd-action}) as in \cite{Regge:1957td, Zerilli:1970se},
\begin{align}
\mathcal{I}_\mathrm{odd}=&\, \frac{j^2 \psi\sqrt{\frac{h_1}{h}}}{2}{\left({\dot{l}_1}-l_0'+\frac{2 {l_0}}r\right)}^2-\frac{j^2\left(\psi\sqrt{\frac{h_1}{h}}
+r\left[\sqrt{\frac{h_1}{h}}\psi\right]'\right) {l_0}^2}{r^2} \nonumber\\
& +\frac{j^2 l_0{}^2}{2r^2}\left[2\left(r\left\{\psi \sqrt{\frac{h_1}{h}}\right\}' +\psi\sqrt{\frac{h_1}{h}}\right)
+\frac{\left(j^2-2\right)\psi}{\sqrt{hh_1}} \right]-\frac{j^2 \left(j^2-2\right) \sqrt{hh_1} \psi l_1^2}{2r^2}\,.
\label{eq:Lodd2}
\end{align}
By using the Lagrange multiplier $\lambda$, Eq.~(\ref{eq:Lodd2}) can be rewritten as,
\begin{align}
\mathcal{I}_\mathrm{odd}=&\, \frac{j^2 \psi\sqrt{\frac{h_1}{h}}}{2}\left[2 \lambda\left(\dot{l}_1-l'_{\mathrm{0}}+\frac{2 l_{0}}{r}\right)-\lambda^2\right]
 -\frac{j^2\left(\psi\sqrt{\frac{h_1}{h}} +r\left[\sqrt{\frac{h_1}{h}}\psi\right]'\right) {l_0}^2}{r^2}\nonumber\\
&+\frac{j^2 l_0^2}{2r^2}\left[2\left(r\left\{\psi \sqrt{\frac{h_1}{h}}\right\}' +\psi\sqrt{\frac{h_1}{h}}\right)
+\frac{\left(j^2-2\right)\psi}{\sqrt{hh_1}} \right]-\frac{j^2 \left(j^2-2\right) \sqrt{hh_1}\, \psi {l_1}^2}{2r^2}\,.
\label{eq:Lodd3}
\end{align}
By the variation of the action corresponding to Eq.~(\ref{eq:Lodd3}) with respect to $l_1$ and $l_1$, we obtain the equations,
which can be solved with respect to $l_1$ and $l_1$ as follows,
\begin{align}
l_1 =&\, -\frac{r^2 \dot{\lambda}}{\left(j^2-2\right)h}\,,
\label{eq:oddh1}\\
l_0 =&\, \frac{r\left[\left\{\psi\sqrt{\frac{h_1}{h}}+2r\left[\sqrt{\frac{h_1}{h}}\psi\right]'\right\}q
+2\lambda'r\psi\sqrt{\frac{h_1}{h}}\right]}{2j^2 \left[\psi\sqrt{\frac{h_1}{h}}+r\left[\sqrt{\frac{h_1}{h}}\psi\right]'
 -\left(r\left\{\psi \sqrt{\frac{h_1}{h}}\right\}' +\psi\sqrt{\frac{h_1}{h}}+\frac{\left(j^2-2\right)\psi}{2\sqrt{hh_1}}\right)\right]}\:.
\label{eq:oddh0}
\end{align}
Eq.~(\ref{eq:oddh1}) tells that the dynamical degree of the freedom of $l_1$ is transferred to $\lambda$ and we can regard $\lambda$ as a dynamical field
instead of $l_1$
Because $l_0$ is also given in terms of the dynamical field $\lambda$ by Eq.~(\ref{eq:oddh0}),
by deleting $l_1$ and $l_0$ by using Eqs.~(\ref{eq:oddh1}) and (\ref{eq:oddh0}) in the action (\ref{eq:Lodd3}), we obtain
\begin{align}
\mathcal{I}_\mathrm{odd}=&\, \frac{j^2r^2\psi\sqrt{\frac{h_1}{h^3}}}{2\left(j^2-2\right)}{\dot\lambda}^2-\frac{j^2h_1\, \psi^2 \,\lambda'^2}{4h\left[\left(r\left\{\psi \sqrt{\frac{h_1}{h}}\right\}'
+\psi\sqrt{\frac{h_1}{h}}+\frac{\left(j^2-2\right)\psi}{2\sqrt{hh_1}}\right)-\psi\sqrt{\frac{h_1}{h}}+r\left[\sqrt{\frac{h_1}{h}}\psi\right]'\right]}-\beta^2 \lambda^2 \nonumber \\
&\, + \left(\mbox{total derivative terms}\right)\, ,
\label{eq:LoddF}
\end{align}
where
\begin{align}
\beta^2=\frac{b_1r^2\left[r^2b'_1b'_3-r^2b''_1b_3+2b_1b_3+4{b'_1}^2+r^2{b_3}^2-2b_1b''_1+2rb_1b'_3-4rb'_1b_3\right]}{\left(2b_1+2rb'_1-r^2b_3\right)^2}\,,
\end{align}
and
\begin{align}
b_1=\frac{j^2\psi \sqrt{h_1}}{2\sqrt{h}}\, ,\quad b_2=\frac{2j^2\psi\left(j^2-2\right)\sqrt{hh_1}}{r^2}\, , \quad
b_3=j^2r^2\left(\frac{\psi \sqrt{h_1}}{\sqrt{h}}+r\left\{\frac{\psi \sqrt{h_1}}{\sqrt{h}}\right\}'+\frac{\left(j^2-2\right)\psi}{2\sqrt{hh_1}}\right)\,.
\end{align}
By the coefficient of ${\dot\lambda}^2$ in the action Eq.~(\ref{eq:LoddF}), we find the condition for the absence of ghosts
\begin{align}
\label{noghost}
j^2\geq2\, ,\quad \mbox{and} \quad \psi \sqrt{\frac{h_1}{h^3}}\geq0\,.
\end{align}
Thus, the solutions for $\lambda$ and therefore $l_1$ and $l_0$, which proportional to $\e^{i(\omega t-kr)}$, when
$k$ and $\omega$ are large, the radial dispersion relation is given as
\begin{align}
\label{dispersion}
\omega^2=h h_1 k^2\, .
\end{align}
Therefore, the radial speed reads
\begin{align}
\label{radialspeed}
c_{\mathrm{odd}}^2=\left(\frac{dr_{*}}{d\tau}\right)^2=1\,,
\end{align}
where $r_*$ is the radial tortoise coordinate, defined by $dr_{*}^2=dr^2/h_1$
and $\tau$ is the proper time, $d\tau^2=h\, dt^2$.

We did not investigate the perturbation for the parity-even modes but as in the standard $F(R)$ gravity, the modes could correspond to the
propagation of the standard spin-two gravitational wave and the spin-zero scalar mode which is specific to the $F(R)$ gravity.

\section{Discussion and conclusions }\label{S77}

Static space-time with spherical symmetry gives an important application for black hole physics \cite{Chakraborty:2016lxo}.
Especially in the case of $-g_{tt}=\frac{1}{g_{rr}}$, by using a specific form of $F(R)$, many spherically symmetric solutions
have been derived \cite{Nashed:2019tuk,Elizalde:2020icc,Nashed:2018oaf,Nashed:2018efg,Nashed:2018piz}.
In this study, we considered a static and spherically symmetric space-time including the case of $-g_{tt}\neq \frac{1}{g_{rr}}$
and we did not assume any specific form of the $F(R)$ gravity theory.

First, we stress the following facts:
\begin{enumerate}
\item We separated the expression of $F(R)$ on one side using the trace of the equation of motions of $F(R)$ with electromagnetic fields.
\item By using Eq.~(\ref{f3ss}), we obtained the equation of motions for $F(R)$ gravity coupled with electromagnetic fields.
The equation involves the first derivative of $F(R)$ concerning the Ricci scalar, $R$, i.e., $F_R=\frac{dF(R)}{dR}$ but does not invole $F$ itself.
By using the equation in the space-time given by Eq.~(\ref{met12}) and with electromagnetic fields,
we derived the non-linear differential equations which controlled this system.
We have solved this system exactly in both cases of $-g_{tt}=\frac{1}{g_{rr}}$ and $-g_{tt}\neq \frac{1}{g_{rr}}$.
\begin{enumerate}
\item When $-g_{tt}=\frac{1}{g_{rr}}$, we have shown that $F_R=\frac{dF(R)}{dR}$ must be a constant, which tells that the Ricci scalar is constant but
the electromagnetic field is non-trivial.
In this case, the solution coincides with that in \cite{Nashed:2019tuk,Elizalde:2020icc,Nashed:2018oaf,Nashed:2018efg,Nashed:2018piz}.
\item For the case $-g_{tt}\neq \frac{1}{g_{rr}}$, by assuming $F_R=\frac{dF(R)}{dR}$ have a specific form, i.e., $F_R=1+\frac a{r^3}$,
we solved the system of the non-linear field equations exactly and obtained a solution for the metric and the electric and magnetic fields.
We have shown that the Ricci scalar is not a constant and by using the form of the obtained Ricci scalar, we found the expression of $F(R)$
in a form of power expansion concerning the Ricci scalar.
The main feature in the case $-g_{tt}\neq \frac{1}{g_{rr}}$ is that the solution cannot reproduce the Reissner-Nordstr{\" o}m metric
of general relativity in any limit.
This means that the obtained black hole solution is a new charged exact solution in $F(R)$ gravity theory.
If the electromagnetic fields and the parameter $a$ vanish, we recover the Schwarzschild space-time.
Due to the complicated forms of the metric, we have considered their asymptotic forms when the radial coordinate $r$ is large,
and we have shown that they asymptotically approach AdS/dS space-time.
In spite that the field equations of $F(R)$ with electromagnetic fields did not involve a cosmological constant,
the metric asymptotically approaches AdS/dS which means that $F(R)$ acts as a cosmological constant.
This effective cosmological constant played an important role in the study of horizons.
We have shown that when the effective cosmological constant has a positive value, we obtain two horizons and
when the effective cosmological constant is negative, we obtain a black hole with three horizons.
In \cite{Jaime:2010kn}, it has been shown that the conditions for the absence of the ghost is given by $\frac{d F(R)}{d R} > 0$ and $\frac{d^2 F(R)}{d R^2} > 0$.
It is important to stress that our black hole solution in Eq.~(\ref{mpabB}) also satisfies the above two conditions.
\end{enumerate}
\end{enumerate}

The Birkhoff theorem has been studied in the frame of the $F(R)$ gravity theory \cite{Riegert:1984zz}.
The problem of the Birkhoff theorem in the $F(R)$ gravity has been studied by several authors trying to explain
if it is valid or not \cite{Sotiriou:2011dz,Sebastiani:2010kv,PerezBergliaffa:2011gj,Gao:2016rdu,Amirabi:2015aya,Calza:2018ohl,Oliva:2011xu,Capozziello:2011wg}.
The present study did not assume any approximation or conformal transformation to obtain the exact black hole solution (\ref{mpabB}) but
the results obtained in this study ensure that the Birkhoff theorem is not true for $F(R)$ gravity theories \cite{Xavier:2020ulw}.
It is known that Birkhoff's theorem is true in general relativity because of the non-existence of zero spin modes in the linear form of the field equations.
When the zero spin mode is absent, the spherically symmetric space-time does not couple to higher-spin excitations \cite{Misner:1973prb, Riegert:1984zz}.
Thus, in the frame of $F(R)$ gravity theory, the differential equation verified by the Ricci scalar, $R$, is the source of zero spin modes.
Therefore, a non-dependence between the Ricci scalar and the metric, in general, yields the non-validity of the Birkhoff theorem in $F(R)$.
This is the case of the exact solution given by Eq.~(\ref{mpabB}) which yields a dynamical value of the Ricci scalar $R$.

Moreover, we investigated the inherent physics of the black hole (\ref{mpabB}) by evaluating its scalar invariants given by the squares of the curvatures
and showed all its behavior up to the leading order
as $\left(R_{\mu \nu \rho \sigma} R^{\mu \nu \rho \sigma},R_{\mu \nu} R^{\mu \nu},R \right)
= \left( \mathcal{O}\left(\frac{1}{r^3}\right), \mathcal{O}\left(\frac{1}{r^3}\right),\mathcal{O}\left(\frac{1}{r^3}\right) \right)$.
These behaviors do not coincide with those in the Reissner-Nordstr{\" o}m black hole which yields the leading order of the Kretschmann scalar
as $\mathcal{O}\left(\frac{1}{r^6}\right)$ and the Ricci tensor squared $R_{\mu \nu} R^{\mu \nu}=\mathcal{O}\left(\frac{1}{r^8}\right)$
and $R=\mathrm{const.}$.
This shows clearly that the singularity of the black hole (\ref{mpabB}) for the Kretschmann scalar is much milder than the black hole of general relativity.
We stress that this merit is generated because of the contribution from the higher-order curvature of $F(R)$, i.e.,
the dimensional parameter $a$.

To clarify the physical properties of the obtained black hole, we calculated some thermodynamical quantities like the entropy, the Hawking temperature,
the quasi-local energy, and the Gibbs energy for the case with the effective positive cosmological constant.
We have shown that all thermodynamical quantities are consistent with what is presented in the previous literature.
Mainly, we have shown that the temperature relies on the radius horizon $r_2$.
When $r_2$ is less than the degenerate horizon, we obtained a negative temperature and the contrary is valid.
We did the same calculations when the effective cosmological constant takes a negative value.
We have found that the degenerate horizons have an essential role to make the temperature takes a positive value.
At the same time, we proved that the black hole verified the first law of thermodynamics.

Another test to examine the black hole (\ref{mpabB}) was the study of its stability.
For this purpose, we rewrote the Lagrangian of the $F(R)$ gravity by using a scalar field that is coupled with the Ricci scalar-tensor.
By investigating the odd-type modes in the perturbation, we have obtained the gradient instability condition and the speed of the radial propagation in the parity-odd mode
of the perturbation and find that the speed is equal to unity, that is the speed of light, for black holes (\ref{mpabB}) and (\ref{mpab1}).

Finally, we stress that the black hole solutions in Eq.~(\ref{mpabB}) and (\ref{mpab1}) are not general solution of the $F(R)$ gravity.
The reason for this is the fact that in this paper, we have supposed that $F=F_R=1+\frac{a}{r^3}$ to obtain the black hole (\ref{mpabB}) and (\ref{mpab1}).
Maybe if we assume another form of $F$, we will obtain another new black hole which will be different from the black hole given by Eq.~(\ref{mpabB}) and (\ref{mpab1}).
{Moreover, there could be any analytic spherically symmetric interior solution can be derived from the field Eq.~(\ref{f3ss}) following the procedure done in \cite{Nashed:2022yfc}.}

\end{document}